\documentclass{article}
\usepackage[utf8]{inputenc}

\usepackage{amsmath,amssymb,bbm,amsthm}
\usepackage{fullpage}
\usepackage{thm-restate,color,xcolor,xspace}
\usepackage{hyperref,cleveref}
\usepackage{algorithm}
\PassOptionsToPackage{noend}{algpseudocode}
\usepackage{algpseudocode}
\usepackage{thm-restate}
\usepackage{graphicx}
\usepackage{tikz}
\usepackage{colortbl}
\usepackage{mathtools}
\usepackage{dsfont}
\usepackage{comment}

\usepackage{pgfplots}
\pgfplotsset{compat=1.17}

\algrenewcommand\algorithmicrequire{\textbf{Input:}}

\usepackage{amsmath}

\usepackage{booktabs}
 \usepackage{multirow} 

 \usepackage{xcolor}

\newtheorem{theorem}{Theorem}[section]
\newtheorem{definition}[theorem]{Definition}
\newtheorem{lemma}[theorem]{Lemma}

\newtheorem*{theorem*}{Theorem}
\newtheorem*{corollary*}{Corollary}
\newtheorem{corollary}[theorem]{Corollary}

\newtheorem{claim}[theorem]{Claim}

\def \eps {\varepsilon}

\definecolor{verylightgray}{gray}{0.9}

\title{Tight Bounds for Low-Error Frequency Moment Estimation and the Power of Multiple Passes}
\author{
Naomi Green-Maimon\\
  \emph{Tel Aviv University}
  \and
Or Zamir\\
  \emph{Tel Aviv University}
}
\date{}
\begin{document}

\maketitle

\begin{abstract}
Estimating the second frequency moment $F_2$ of a data stream up to a $(1 \pm \varepsilon)$ factor is a central problem in the streaming literature. For errors $\varepsilon > \Omega(1/\sqrt{n})$, the tight bound $\Theta\left(\log(\varepsilon^2 n)/\varepsilon^2\right)$ was recently established by Braverman and Zamir. In this work, we complete the picture by resolving the remaining regime of small error, $\varepsilon < 1/\sqrt{n}$, showing that the optimal space complexity is $\Theta\left( \min\left(n, \frac{1}{\varepsilon^2} \right) \cdot \left(1 + \left| \log(\varepsilon^2 n) \right| \right) \right)$ 
bits for all $\varepsilon \geq 1/n^2$, assuming a sufficiently large universe. This closes the gap between the best known $\Omega(n)$ lower bound and the straightforward $O(n \log n)$ upper bound in that range, and shows that essentially storing the entire stream is necessary for high-precision estimation.

To derive this bound, we fully characterize the two-party communication complexity of estimating the size of a set intersection up to an arbitrary additive error $\varepsilon n$. 
In particular, we prove a tight $\Omega(n \log n)$ lower bound for one-way communication protocols when $\varepsilon < n^{-1/2-\Omega(1)}$, in contrast to classical $O(n)$-bit protocols that use two-way communication.
Motivated by this separation, we present a two-pass streaming algorithm that computes the exact histogram of a stream with high probability using only $O(n \log \log n)$ bits of space, in contrast to the $\Theta(n \log n)$ bits required in one pass even to approximate $F_2$ with small error. This yields the first asymptotic separation between one-pass and $O(1)$-passes space complexity for small frequency moment estimation.

\end{abstract}

\section{Introduction}
The streaming model of computation is fundamental for processing massive datasets and has been extensively studied in both theory and practice. A central challenge in this model is the \emph{frequency moment estimation} problem, in which an algorithm observes a sequence of elements from a universe~$U$ and must approximate a function of their frequency distribution using limited memory.
Denote by~$f_x \in \mathbb{N}$ the number of times an element~$x\in U$ appeared in the stream; the algorithm aims to return, with probability higher than~$99\%$, a~$(1\pm \varepsilon)$-estimation of~$F_p := \sum_{x\in U} f_x^p$ --- the \emph{$p$-th frequency moment} of the stream. We generally denote the length of the stream by~$n$ and assume that~$|U|=\text{poly} (n)$.
In fact, we can usually assume that~$|U|\leq O(n^2)$ as hashing the elements into a set of size~$\Theta(n^2)$ is expected to not change the frequencies at all. 
The primary resource constraint studied in this setting is the amount of space required for a successful approximation. This line of research originated with the foundational work of Alon, Matias, and Szegedy~\cite{alon1996space}, which both introduced this problem and established the first space-efficient algorithms for frequency moment estimation.

Among the frequency moments, the case of~$p=2$, known as \emph{second moment estimation}, has particular significance. This quantity, sometimes referred to as the \emph{repeat rate} or \emph{surprise index}, plays a key role in applications such as database query optimization~\cite{alon1999tracking}, network traffic anomaly detection~\cite{krishnamurthy2003sketch}, and approximate histogram maintenance~\cite{gilbert2002fast}.
For~$F_2$, the original algorithm proposed by Alon, Matias, and Szegedy achieves a space complexity of~$O(\log n / \varepsilon^2)$ bits.
Only recently, Braverman and Zamir~\cite{braverman2025optimality} obtained a matching lower bound for a wide range of error parameters~$\varepsilon$. 
They showed that for any~$\varepsilon > \Omega(1/\sqrt{n})$ the optimal space complexity for estimating the second frequency moment is $\Theta(\log(\varepsilon^2 n)/\varepsilon^2)$.
This leaves the range of~$\varepsilon<1/\sqrt{n}$ as the only range without a tight space bound for the problem. 
In that range, the best known lower bound is~$\Omega(n)$ which follows from either~\cite{braverman2025optimality} or the much earlier~\cite{woodruff2004optimal} for~$\varepsilon=1/\sqrt{n}$.
On the other hand, by simply maintaining the entire histogram of the input stream in memory we obtain an upper bound of~$O\left(n \log \left(2 + \frac{|U|}{n}\right)\right) = O(n\log n)$ even for \emph{exactly} computing the second frequency moment.
We note that it suffices to consider only~$\varepsilon \geq 1/n^2$ as the second frequency moment is a natural number in the range~$[n,n^2]$.
We are thus left with a gap between the~$\Omega(n)$ lower bound and~$O(n\log n)$ upper bound in the range of small error~$1/n^2 \leq \varepsilon \leq 1/\sqrt{n}$.

In this work, we establish the space complexity of second moment estimation in this remaining range of error parameters. 

\begin{theorem}\label{thm:f2lb}
    The space complexity of $\left(1\pm \varepsilon\right)$-estimating the second frequency moment of a stream of~$n$ elements, for any~$\frac{1}{n^2}\leq \varepsilon \leq \frac{1}{\sqrt{n}}$, is~$$
    \Theta\left(n \log \left(\frac{1}{\varepsilon^2 n}\right)\right)
    .$$
\end{theorem}

\begin{figure}
\centering
\begin{tikzpicture}
    \begin{axis}[
        xlabel={Error ($\varepsilon$)},
        ylabel={Space},
        xtick={0.345,1},
        xticklabels={\small $1/\sqrt{n}$,},
        yticklabels={},
        ymin=0, ymax=5,
        xmin=0, xmax=1,
        width=12cm, height=7cm,
        grid=major,
        axis lines=middle,
        every axis x label/.style={at={(ticklabel* cs:1.05)}, anchor=north},
        every axis y label/.style={at={(ticklabel* cs:1.05)}, anchor=east}
    ]

    \addplot[domain=0:0.33, samples=50, thick, black] {4} node[pos=0.5, above] { $\Theta(n \log n)$};
    
    \addplot[only marks, mark=*, thick] coordinates {(0.33, 4)}; 
    \addplot[only marks, mark=*, thick] coordinates {(0.36, 3)}; 
    \draw[dashed] (axis cs:0.33,0) -- (axis cs:0.33,5) node[midway, right] {\small };
    \draw[dashed] (axis cs:0.36,0) -- (axis cs:0.36,5) node[midway, right] {\small };
    
    \addplot[domain=0.33:0.36, samples=50, thick, gray] {-148148.15*x^3 + 153333.33*x^2 - 52900*x + 6087} node[pos=0.25, below right] { $\Theta(n)$};
    
    \addplot[domain=0.36:1, samples=50, thick, black] {0.389/x^2} 
    node[pos=0.8, above] { $\Theta\left(\frac{\log n}{\varepsilon^2}\right)$};

    \end{axis}
\end{tikzpicture}
\caption{The optimal space complexity of second moment estimation for all values of~$\varepsilon$.}
\label{fig:plot} 
\end{figure}
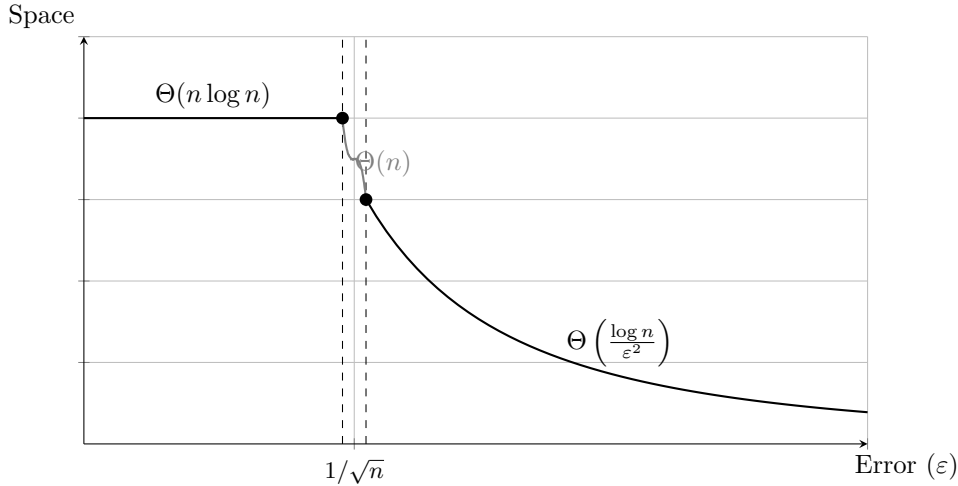

We observe that this bound implies a phase transition around~$\varepsilon = 1/\sqrt{n}$: if the wanted error is polynomially-smaller than~$1/\sqrt{n}$ then the space required is the same as needed to maintain the entire input in memory and compute an \emph{exact} answer, while if~$\varepsilon = 1/\sqrt{n}$ then only~$\Theta(n)$ bits of space are needed. 
Our bounds combined with those of~\cite{braverman2025optimality} give a complete characterization of the space complexity needed to estimate the second moment up to any error parameter~$\varepsilon$. See Figure~\ref{fig:plot} for a graph. Note that both lower bounds assume that the universe of possible stream elements is sufficiently large.

To derive our lower bound, we study and fully characterize the two-party communication complexity of the \emph{Approximate Set Intersection} problem.
In this problem, Alice and Bob are given subsets~$A$ and~$B$ of a larger universe~$U$, of size at most~$n$ each, and need to output~$|A\cap B|$ up to an additive error~$\varepsilon n$ with probability at least~$2/3$.
This is a natural generalization of several classical problems in communication complexity: \emph{Set Disjointness} is a special case of the problem when~$\varepsilon=0$, and of \emph{Gap Hamming Distance} when~$\varepsilon = 1/\sqrt{n}$. 
When~$\varepsilon < 1/\sqrt{n}$ a lower bound of~$\Omega(n)$ follows from~\cite{woodruff2004optimal} and a matching upper bound of~$O(n)$ holds for exactly computing the intersection with two-way communication, even when the element universe~$U$ is allowed to be very large~\cite{brody2014}.
To the best of our knowledge, no tight bounds were explicitly derived in all other settings and ranges of parameters. 
In~\cite{Meir2018} the same problem was asked and discussed, resulting only in the rather straightforward~$O(\log n / \varepsilon^2)$ upper bound obtained by Alice explicitly sending a sample of her elements to Bob. 
We obtain the following optimal bounds.

\begin{theorem}\label{thm:approxset}
    For~$\varepsilon<1/\sqrt{n}$, the one-way communication complexity of Approximate Set Intersection is~$\Theta(n \log(2/\varepsilon^2 n))$ and the two-way complexity is~$\Theta(n)$.
    For~$\varepsilon\geq 1/\sqrt{n}$ the communication complexity of Approximate Set Intersection is~$\Theta(1/\varepsilon^2 +\log n)$ in both the one-way or two-way settings.
\end{theorem}

The gap between the optimal bounds for Approximate Set Intersection in the one-way and two-way communication settings implies that our lower bound for~$F_2$ estimation in the range~$\varepsilon<1/\sqrt{n}$ does not extend to multi-pass streaming algorithms. 
Motivated by that, we present a streaming algorithm extending the ideas from the communication protocol of Brody et al.~\cite{brody2014}, and computes an exact histogram of a stream (and in particular its exact~$F_2$) using only two passes and~$O(n\log\log n)=o(n\log n)$ memory.

\begin{theorem}\label{thm:multipass}
    There exists a streaming algorithm that in two passes and using~$O(n\log\log n)$ bits of memory computes, with high probability, an exact histogram of the stream.
\end{theorem}

Woodruff and Zhou~\cite{woodruff2021separations} recently showed that for~$p>2$ and~$\varepsilon=o(1)$, there is an asymptotic separation between the space complexity of~$F_p$ estimation in one-pass versus~$O(1)$-passes. They left as an open problem the question of whether there is such a separation also for ``small" frequency moments (that is,~$p\leq2$).
We show that indeed there exists such separation for~$F_2$, at least in the range~$\varepsilon<1/\sqrt{n}$.
We then generalize this streaming algorithm for a larger number of passes and conclude that for exact histogram computation with $r$ passes, $\Theta\!\bigl(n \log^{(\Theta(r))}\! n\bigr)$ bits of memory are both necessary and sufficient.

\begin{figure}
\centering
\begin{tabular}{@{}llcc@{}}
\toprule
\textbf{Problem} & \textbf{Model} & $\varepsilon < 1/\sqrt{n}$ & $\varepsilon \ge 1/\sqrt{n}$ \\
\midrule
\multirow{2}{*}{Approx. Set Intersection} 
  & One-way & \boldmath{$\Theta(n \log(2/\varepsilon^2 n))$} & \boldmath{$\Theta(1/\varepsilon^2+\log n)$} \\
  & Two-way & $\Theta(n)$ & \boldmath{$\Theta(1/\varepsilon^2+\log n)$} \\
\midrule
\multirow{2}{*}{$F_2$ Estimation} 
  & One-pass & \boldmath{$\Theta(n \log(2/\varepsilon^2 n))$} & $\Theta(\log(\varepsilon^2 n)/\varepsilon^2)$ \\
  & $O(1)$-passes & \boldmath{$O(n \log \log n)$} & 
    $\begin{array}{l}
      O(\log(\varepsilon^2 n)/\varepsilon^2), 
      \Omega(1/\varepsilon^2+\log n)
     \end{array}$ \\
\bottomrule
\end{tabular}
\caption{Complexity of Approximate Set Intersection and $F_2$ Estimation.}
\label{fig:table}
\end{figure}

See Figure~\ref{fig:table} for a table summarizing the best-known bounds for the communication and streaming problem we discuss, with results in bold being those presented in this paper.

The space complexity of frequency moment estimation was extensively studied in many settings, and for many parameter choices~\cite{chakrabarti2003near,woodruff2004optimal,bar2004information,indyk2005optimal,bhuvanagiri2006simpler,monemizadeh20101,braverman2010recursive,kane2010exact,ganguly2011lower,ganguly2011polynomial,andoni2011streaming,woodruff2012tight,li2013tight,andoni2017high,braverman2018revisiting,jayaram2019towards,nelson2022optimal}, we refer the reader to the introduction of~\cite{braverman2025optimality} for a more exhaustive overview of the existing literature.

\section{Overview and Organization}
In this section, we outline the main technical ideas behind our results and describe the organization of the rest of the paper.

\paragraph{Section 3: Communication lower bounds and streaming implications, in the~$\varepsilon<1/\sqrt{n}$ range.} We prove that any one-way communication protocol that approximates $|A \cap B|$ up to additive error $\varepsilon n$ requires $\Omega(n \log(1/\varepsilon^2 n))$ bits when $\varepsilon < 1/\sqrt{n}$. This is shown via an information-theoretic construction inspired by the construction used by~\cite{dasgupta2012} to obtain a lower bound for the one-way communication complexity of Set Disjointness. 
We construct a large family of inputs for Alice and a small set of inputs for Bob, and show that every Alice input is uniquely identified by the answers of the communication protocol to all of Bob's possible inputs. 
In a one-way communication protocol though, Alice's message depends only on her own input and should thus intuitively encode enough information to answer any possible Bob input --- which in this case results in having to fully encode her own input.
Our main contribution is proving the existence of such input families, which becomes more involved in comparison to the exact problem of Set Disjointness. 
We then reduce from this communication problem to streaming $F_2$ estimation using a standard concatenation protocol, implying the same lower bound for streaming algorithms.

\paragraph{Section 4: Matching upper bounds via hashing.} We complement the lower bounds with a simple estimator for $F_2$ that matches the bound up to constants. The algorithm hashes the universe into a domain of size $m = \Theta(1/\varepsilon^2)$ using a 4-wise independent hash function, and stores the exact multiset of hash values. 
Here the main observation is that while the hashing induces too many collisions, we can analytically subtract the expected number of hash collisions from the observed number of collisions to obtain a significantly better estimator for the \emph{number} of collisions in the original stream.
We analyze the variance of the estimator and show it concentrates around the true $F_2$ value, yielding an estimate with additive error $\varepsilon F_2$ using $O(n \log(1/\varepsilon^2 n))$ bits of memory.

\paragraph{Section 5: The high-error regime.} We analyze the case $\varepsilon \geq 1/\sqrt{n}$ and prove that the communication complexity of approximate set intersection is $\Theta(1/\varepsilon^2 + \log n)$ in both one-way and two-way models. 
Both the lower and upper bounds are essentially reductions to the~$\varepsilon=\Theta(1/\sqrt{n})$ case which is very similar to the Gap Hamming Distance Problem.
The lower bound is then indeed established by a reduction from the Gap Hamming Distance problem. 
The matching upper bound is obtained using subsampling which reduces the problem to the problem of estimating the set intersection size of two sets of some size~$k$ up to an additive error of~$\Theta(\sqrt{k})$, a problem which we already present an optimal solution for in the previous sections.
These simple bounds complete our characterization of the communication complexity of the Approximate Set Intersection problem.

\paragraph{Section 6: A two-pass algorithm for exact histograms.} Motivated by the one-way/two-way gap in communication, we begin by designing a streaming algorithm that computes the exact histogram of the stream in three passes using only $O(n \log \log n)$ bits of space. The algorithm uses the first pass to hash elements into buckets and record “cheap” fingerprint-based histograms in each bucket, then uses the second pass to verify which of these histograms is correct using a randomized algebraic identity check, and finally uses the third pass to recompute the histograms of the few incorrect buckets exactly. 
Our algorithm builds upon the communication protocol of Brody et al.~\cite{brody2014} for computing the exact set intersection in two-way communication, and transforms it to a streaming algorithm. One of our main technical contributions in this algorithm is constructing a streaming algorithm that can verify, in one-pass, an assumed histogram for a multiset of elements. This verification step is straightforward in the two-way communication setting, where each party can hash and compare their entire input, but requires new techniques in the streaming model.
We then introduce ideas from sparse recovery to eliminate the third pass and achieve the same memory bound with a two-pass streaming algorithm.
Finally, we discuss the optimality of this bound, concluding that any constant-pass algorithm for computing even the~$F_2$ requires~$\Omega(n \log^{(r)}n)$ memory for some constant~$t$. In Appendix~\ref{app:rpass} we generalize our algorithm to a larger number of passes.

\paragraph{Section 7:} We conclude with a summary of our results and a discussion of open problems. In particular, we highlight the question of whether a separation between one-pass and multi-pass complexity exists also in the lower-accuracy regime $\varepsilon > 1/\sqrt{n}$.

\section{Lower Bounds in the Low Error Regime}
In this section, we prove the lower bounds in Theorems \ref{thm:f2lb} and \ref{thm:approxset} in the $\eps<1/\sqrt{n}$ regime. We first establish the communication complexity lower bound for computing set intersection size up to small additive error, and then show this result implies a lower bound for second-moment estimation in the streaming model under similar error guarantees.
Throughout this section, we assume the universe $U$ is of large enough (yet polynomial) size in $n$.

\subsection{Approximate Set Intersection}
Our proof follows the general framework of Dasgupta et al.~\cite{dasgupta2012}, who proved a~$\Omega(n \log n)$ lower bound for the one-way communication complexity of Set Disjointness.
We begin by defining a promise problem that captures the essential hardness of approximate set intersection, which can be viewed as a refinement of the Gap Hamming Distance promise problem.
We then construct, non explicitly, a large family of inputs for Alice and a much smaller family for Bob such that, for every input in Alice’s family, the answers to the promise problem for all inputs in Bob’s family uniquely identifies it. 
Such a family implies a lower bound on the one-way communication complexity. 

\begin{definition}[$\mathrm{INT}^{\varepsilon}$]
Let $X,Y\subset U$ inputs for Alice and Bob respectively, such that $|X|,|Y|\le n$. 
We denote by $\mathrm{INT}^{\varepsilon}(X,Y)$ the communication problem of calculating $|X \cap Y|$ up to an additive error $\varepsilon n$ with probability $>2/3$.
\end{definition}

\begin{definition}[$\text{INT-P}^\alpha$ Promise Problem]
Let $\alpha=\alpha(n)\in \left(\frac{1000}{\ln n},\frac{1}{2}\right)$, and $X,Y\subset U$ inputs for Alice and Bob respectively, such that $|X|,|Y|\le n$. \\
We denote by $\emph{INT-P}^\alpha$ the communication problem of distinguishing with probability $>2/3$ between
\begin{itemize}
    \item \textbf{Yes-instances:} Pairs $(X,Y)$ such that $|X \cap Y| \ge n^{1 - \alpha} + \frac{1}{2}\sqrt{n^{1 - \alpha}}$.
    \item \textbf{No-instances:} Pairs $(X,Y)$ such that $|X \cap Y| \le n^{1 - \alpha} - \frac{1}{2}\sqrt{n^{1 - \alpha}}$.
\end{itemize}
For brevity, we write $(X,Y)\in\textsc{Yes}$ or $(X,Y)\in\textsc{No}$, always referring to this promise problem.
\end{definition}

We observe that any protocol solving~$\mathrm{INT}^\varepsilon$ also solves~$\mathrm{INT-P}^\alpha$ as long as~$\varepsilon n < \frac{1}{2}\sqrt{n^{1-\alpha}}$, or equivalently, $\alpha <\frac{\log\left({1}/\left({\varepsilon^2 n}\right)\right) - 2}{\log n}$.
We thus first establish an~$\Omega(\alpha n \log n)$ lower bound for~$\mathrm{INT-P}^\alpha$ and then deduce from it an~$\Omega\left(n\log\left({1}/\left({\varepsilon^2 n}\right)\right)\right)$ lower bound for~$\mathrm{INT}^\varepsilon$.
We begin by constructing a large family~$\mathcal{X}$ of possible inputs to Alice, and a small family~$\mathcal{Y}$ for possible inputs to Bob, such that the answers to~$\mathrm{INT-P}^\alpha$ on all inputs of~$\mathcal{Y}$ uniquely identifies any input~$X\in\mathcal{X}$.

\begin{definition}[Design]
A family $\mathcal{X}$ of subsets of $U$ is a \emph{design} if:
\begin{itemize}
    \item $\forall X \in \mathcal{X}$, we have $|X| = n$,
    \item $\forall X \ne X' \in \mathcal{X}$, we have $|X \cap X'| \le \frac{n}{400}$.
\end{itemize}
\end{definition}

\begin{definition}[Block-Uniform Distribution $\mathcal{D}_{n,\alpha}$]
  Fix an integer $n\ge 1$ and a parameter $\alpha=\alpha(n)\in(\frac{1000}{\ln n},1)$.
  Partition the universe $[n^{1+\alpha}]\subseteq U$ into $n$ disjoint
  blocks $B_1,\dots,B_n$, each of size $n^{\alpha}$.
  The \emph{block-uniform distribution} $\mathcal{D}_{n,\alpha}$ is the
  uniform distribution over sets $X\subset[n^{1+\alpha}],|X|=n$ obtained by the
  following sampling procedure:
  \begin{enumerate}
    \item For each block $B_i$ independently, pick one element
          $x_i\in B_i$ uniformly at random.
    \item Output the set $X=\{x_1,x_2,\dots,x_n\}$.
  \end{enumerate}
\end{definition}

\begin{lemma}\label{lem:design-exists}
For a sufficiently large integer $n$ and any $\alpha = \alpha(n) \in (\frac{1000}{\ln n},\frac{1}{2})$, there exists a design of size $\exp(c \cdot \alpha n \ln n)$ for some constant $c$.
\end{lemma}

\begin{proof}
We give a non-explicit construction using the probabilistic method. Independently and uniformly sample $N:=\exp\left(\frac{1}{3200}\alpha n \ln n\right)=\exp\left(c\cdot \alpha n\ln n\right)$ sets from $\mathcal{D}_{n,\alpha}$.

Let $X, X'$ be two sets sampled from~$\mathcal{D}_{n,\alpha}$. Let $Y\coloneqq |X\cap X'|$, and note that $Y \sim \text{Bin}(n, \frac{1}{n^\alpha})$ and $\mu \coloneqq \mathbb{E}[Y] = n \cdot \frac{1}{n^\alpha} = n^{1-\alpha}$. Using a standard Chernoff bound we have
\[
\Pr[Y \ge (1 + \delta)\mu] \le \left( \frac{e^{\delta}}{(1 + \delta)^{1 + \delta}} \right)^{\mu},
\]
with $(1 + \delta)\mu = \frac{n}{400}$, so $\delta = \frac{1}{400} n^\alpha - 1$. Note that as $\alpha\ge\frac{1000}{\ln n}$, $\delta>0$ and the Chernoff bound can be applied. Thus, we deduce that
\[
\ln \Pr\left[Y > \frac{n}{400} \right] \le
\mu \left( \delta - (1 + \delta) \ln(1 + \delta) \right) \le
-\frac{1}{2}\mu\cdot\delta \ln(\delta) ,
\]
where the last inequality follows as for all $\delta > 0$, which holds for all $\alpha$ when $n$ is sufficiently large, we have
\[
\delta - (1 + \delta) \ln(1 + \delta) \le -\frac{\delta \ln \delta}{2}.
\]
Note that $\mu\delta
      = n^{1-\alpha} \bigl(\frac{n^{\alpha}}{400}-1\bigr)
      \ge \frac{n}{600}$ and
$\ln\delta\ge\ln\bigl(\frac{n^{\alpha}}{600}\bigr)
             =\alpha\ln n-\ln800$.
The constants $600$ and $800$ are arbitrary values chosen to simplify the expressions and ensure the bound holds for large enough $n$. Hence, for all sufficiently large $n$,
\[
  \Pr \left[Y>\frac n{400}\right]
  \le
  \exp \left(-\frac{\alpha n\ln n}{1600}\right)
.
\]
We apply a union bound over all $\binom{N}{2}$ pairs of sets we sampled, and deduce that the probability that any pair had too big of an intersection is bounded by
\[
\binom{N}{2} \cdot \exp\left( -\frac{1}{1600} \alpha n \ln n \right) < 1.
\]
Therefore, a design of the required size exists for $c=1/3200$.
\end{proof}

\begin{lemma}[Random $Y$ distinguishes between $X$ and $X'$] \label{lem:Y-distinguishes-pair}
There exists a universal constant $p_0 > 0$ such that the following holds.
Let $\alpha = \alpha(n) \in (\frac{1000}{\ln n},\frac{1}{2})$ and a design $\mathcal{X}$, let $X \ne X' \in \mathcal{X}$ be any pair from the design. Then, with $Y$ sampled uniformly from $\mathcal{D}_{n,\alpha}$, 
\[
\Pr_Y\left[ ((X,Y) \in \textsc{Yes} \wedge (X',Y) \in \textsc{No}) \vee ((X,Y) \in \textsc{No} \wedge (X',Y) \in \textsc{Yes}) \right] \ge p_0.
\]
We call this event ``$Y$ distinguishes between $X$ and $X'$".
\end{lemma}

\begin{proof}
Let $p = n^{-\alpha}$ and $\mu = n^{1 - \alpha}$.
Denote by $m = |X \cap X'| \le \frac{n}{400}$ and $M = n - m \ge \frac{399n}{400}$, and denote $\mu_M = Mp, \mu_m = mp$. 
We also denote
\[
Z_{\text{same}} = |Y \cap X \cap X'|,\quad Z_{\text{diff}} = |Y \cap (X \setminus X')|,\quad Z'_{\text{diff}} = |Y \cap (X' \setminus X)|.
\]
\[
Z = |Y \cap X| = Z_{\text{same}} + Z_{\text{diff}},\quad Z' = |Y \cap X'| = Z_{\text{same}} + Z'_{\text{diff}}.
\]
Note that as $X,X',m,M$ are fixed (and are not random variables), and all probabilities are taken only over the random choice of $Y$.
Because the $m$ intersecting blocks and the $M$ differing blocks are disjoint and the choices of $Y$ values in each block are i.i.d, $Z_{\text{same}}$ is independent of $(Z_{\text{diff}},Z'_{\text{diff}})$.

\paragraph{Step 1: Chernoff bound for $Z_{\text{same}}$.}
We have $Z_{\text{same}} \sim \text{Bin}(m, p)$ with mean $\mu_m \le \mu/400$. Using a two-sided Chernoff bound with $\delta = \frac{2}{\sqrt{\mu_m}}$ we have
\[
\Pr[|Z_{\text{same}} - \mu_m| > 2\sqrt{\mu_m}] \le 2e^{-4/3}, \quad\text{so}\quad \Pr[|Z_{\text{same}} - \mu_m| \le 2\sqrt{\mu_m}] \ge c_2 \coloneqq 1 - 2e^{-4/3} > 0.47.
\]

\paragraph{Step 2: Berry–Esseen tail bound for $Z_{\mathrm{diff}}$ and $Z'_{\mathrm{diff}}$.}

Fix an arbitrary ordering of the~$M$ “differing’’ blocks and write  
\[
B_i=\mathds{1}\{Y\cap X'\text{ is not empty in block }i\} \quad i=1,\dots,M.  
\]
Then $B_i\stackrel{\text{i.i.d.}}{\sim}\mathrm{Ber}(p)$ and  
$Z'_{\mathrm{diff}}=\sum_{i=1}^{M}B_i$.  
Center the variables by setting $V_i:=B_i-p$ and  
$S_M:=\sum_{i=1}^{M}V_i$, so that $Z'_{\mathrm{diff}}=\mu_M+S_M$.

\smallskip
\textit{Verifying the Berry–Esseen assumptions.}  
Each $V_i$ satisfies  
$\mathbb{E}[V_i]=0$,  
$\sigma^2:=\operatorname{Var}(V_i)=p(1-p)>0$, and  
$\rho:=\mathbb{E}[|V_i|^{3}]=p\cdot (1-p)^3 + (1-p)\cdot p^3 = p(1-p)\left((1-p)^2+p^{2}\right)\le p(1-p)=\sigma^2$.

\smallskip
\textit{Normalisation.}  
Let $\sigma_M^2=\operatorname{Var}(S_M)=Mp(1-p)=\mu_M(1-p)$ and denote by  
$\widehat S_M:=S_M/\sigma_M$ which has cumulative distribution function  
$F_M(x)=\Pr[\widehat S_M\le x]$.

\smallskip
\textit{Berry–Esseen inequality.}  
By the Berry-Esseen Inequality \cite[Ch.~XVI, §5, Thm.~1]{Feller1971} with the sharpened constant of Shevtsova $C=0.4748$ \cite{Shevtsova2010}, where $\Phi$ denotes the standard normal CDF, we have
\[
|F_M(x)-\Phi(x)|\le
\frac{C\rho}{\sigma^{3}\sqrt{M}}=
\frac{C}{\sqrt{Mp(1-p)}}=
\frac{C}{\sqrt{\mu_M(1-p)}}\le
\frac{\sqrt{2}C}{\sqrt{\mu_M}},
\tag{BE}
\]
where the last inequality holds as $1-p>1/2$.

\smallskip
\textit{Applying the bound.}  
We want a lower bound for $\Pr[Z'_{\mathrm{diff}}\le\mu_M-0.603\sqrt{\mu_M}]$.
Put $k=\mu_M-0.603\sqrt{\mu_M}$ and  
$x_*=(k-\mu_M)/\sigma_M=-0.603/\sqrt{1-p}\in(-0.853,-0.603)$.
Because $\Phi$ is decreasing,
$\Phi(x_*)\ge\Phi(-0.603)=0.2736$.
Applying (BE) and assuming that~$n$ is large enough so $\mu_M\ge100$,
\[
\Pr\bigl[Z'_{\mathrm{diff}}\le\mu_M-0.603\sqrt{\mu_M}\bigr]
   =F_M(x_*)\ge
   \Phi(x_*)-\frac{\sqrt{2}\cdot0.4748}{\sqrt{\mu_M}}
   \ge 0.2736-0.067>0.2 .
\]
Hence,
\[
\boxed{
  \Pr\left[Z'_{\mathrm{diff}}\le\mu_M-0.603\sqrt{\mu_M}\right]\ge c_1:=0.2.
}
\]
By symmetry the same constant bound holds for  
$\Pr\bigl[Z_{\mathrm{diff}}\ge\mu_M+0.603\sqrt{\mu_M}\bigr]$.

\paragraph{Step 3: Combining bounds}
When all three of the above events hold, we get
\[
Z \ge \mu_M + 0.603\sqrt{\mu_M} + Z_{\text{same}} \ge \mu + 0.5\sqrt{\mu},\quad
Z' \le \mu_M - 0.603\sqrt{\mu_M} + Z_{\text{same}} \le \mu - 0.5\sqrt{\mu}.
\]
Which implies
\begin{align*}
    &\Pr\left[ (Z \ge \mu + 0.5\sqrt{\mu}) \wedge (Z' \le \mu - 0.5\sqrt{\mu}) \right] \ge \\
    &\Pr\Big[
        (Z_{\text{diff}} \ge \mu_M + 0.603\sqrt{\mu_M})
        \wedge
        (Z'_{\text{diff}} \le \mu_M - 0.603\sqrt{\mu_M}) \wedge
        (Z_{\text{same}} \in \left[\mu_m - 2\sqrt{\mu_m}, \mu_m + 2\sqrt{\mu_m}) \right]
    \Big] \\
    &= \Pr\left[ (Z_{\text{diff}} \ge \mu_M + 0.603\sqrt{\mu_M})
        \wedge
        (Z'_{\text{diff}} \le \mu_M - 0.603\sqrt{\mu_M}) \right]
    \cdot \Pr\left[Z_{\text{same}} \in \left[\mu_m - 2\sqrt{\mu_m}, \mu_m + 2\sqrt{\mu_m}) \right] \right],    
\end{align*}
where the last equality holds as $Z_{same}$ is independent from $Z_{diff}$ and $Z'_{diff}$.

\paragraph{Step 4: Negative correlation between $Z_{diff}$ and $Z'_{diff}$.}
For every differing index $i$ define the indicators
$E_i=\mathds {1}\{Y_i=X_i\}$ and
$F_i=\mathds {1}\{Y_i\neq X'_i\}$.
Because the random choice of $Y$ is block-wise independent, the
collection $\{(E_i,F_i)\}_{i=1}^{M}$ is also independent.
Consider the two increasing events
\[
  A := \Bigl\{\sum_i E_i\ge\mu_M+0.603\sqrt{\mu_M}\Bigr\},\qquad
  B := \Bigl\{\sum_i (1-F_i)\le\mu_M-0.603\sqrt{\mu_M}\Bigr\}.
\]
Since $E_i=1$ implies $F_i=1$, both $A$ and $B$ are monotone
in the same direction.  By the Harris (binary FKG) inequality
\cite{Harris1960} we thus have
\[
  \Pr[A\wedge B]\ge\Pr[A]\Pr[B].
\]
Rewriting $A$ and $B$ in terms of
$Z_{\mathrm{diff}}$ and $Z'_{\mathrm{diff}}$ gives
\begin{align*}
  &\Pr\bigl[Z_{\mathrm{diff}} \ge \mu_M+0.603\sqrt{\mu_M}
       \wedge
       Z'_{\mathrm{diff}} \le \mu_M-0.603\sqrt{\mu_M}\bigr]\\
  &\ge
  \Pr[Z_{\mathrm{diff}} \ge \mu_M+0.603\sqrt{\mu_M}] \cdot
  \Pr[Z'_{\mathrm{diff}} \le \mu_M-0.603\sqrt{\mu_M}]\ge c_1^2.
\end{align*}

\paragraph{Step 5: Symmetry}
Swapping $X$ and $X'$ gives a disjoint event of equal probability. So the total distinguishing probability is at least
\[
\Pr[Y \text{ distinguishes } X, X'] \ge 2 c_1^2 c_2 =:p_0 > 0.
\]
\end{proof}

\begin{lemma}[Existence of family $\mathcal{Y}$ fully distinguishing a design $\mathcal{X}$] \label{lem:Y-family-describes-X-family}
Let $\alpha = \alpha(n) \in (\frac{1000}{\ln n},\frac{1}{2})$, and let $\mathcal{X}$ be a design of size $\exp(c \cdot \alpha n \ln n)$. There exists a family $\mathcal{Y} \subseteq U$ such that:
\begin{itemize}
    \item $|Y| = n$ for all $Y \in \mathcal{Y}$.
    \item $|\mathcal{Y}| \le k\cdot \alpha n \ln n$ for a universal constant $k$.
    \item For any $X \ne X' \in \mathcal{X}$, some $Y\in \mathcal{Y}$ distinguishes them.
\end{itemize}
We call this behavior $\mathcal{Y}$ describes $\mathcal{X}$.
\end{lemma}

\begin{proof}
We draw a set $\mathcal{Y}$ of the desired size uniformly and independently from $\mathcal{D}_{n,\alpha}$.
Note that the amount of pairs $X \ne X' \in \mathcal{X} $ is smaller than $ \exp(c \cdot \alpha n \ln n)^2 = \exp(2c \cdot \alpha n \ln n)$. As described in Lemma \ref{lem:Y-distinguishes-pair}, for some $X\ne X'$, the probability that a single $y \in \mathcal{Y}$ distinguishes between $X$ and $X'$ is at least $p_0$, (where $p_0$ is the constant probability bound from Lemma \ref{lem:Y-distinguishes-pair}). 
As the sets in $\mathcal{Y}$ are chosen independently, 
\[
\mathbb{P} \left[ \forall Y \in \mathcal{Y}, \text{$Y$ does not distinguish between $X$ and $X'$} \right] < (1-p_0)^{|\mathcal{Y}|}.
\]
Therefore, by a union bound, 
\[
\mathbb{P} \left[ \exists X \ne X' \in \mathcal{X} : \forall Y \in \mathcal{Y}, \text{$Y$ does not distinguish between $X$ and $X'$} \right] < \exp(2c \cdot \alpha n \ln n) \cdot (1-p_0)^{|\mathcal{Y}|},
\]
and for $ |\mathcal{Y}| > \underbrace{-\frac{2c}{\ln (1-p_0)}}_{k} \cdot \alpha n \ln{n}$, this probability is smaller than 1. Meaning, when randomly choosing $\mathcal{Y}$ of size $k\alpha\cdot n \ln{n} $, the probability that it differentiates between every pair $ X\ne X' \in \mathcal{X}$ is positive. Therefore, by the probabilistic method, a family $\mathcal{Y}$ that meets the conditions exist.
\end{proof}

We are now ready to prove the desired lower bound. 
\begin{lemma}\label{thm:lower-bound-INT-low-error}
For any large enough $n$, and for any $\varepsilon<1/(2e^{500}\sqrt n)$, the randomized one-way communication complexity of $\emph{INT}^\varepsilon$ is $\Omega(n\log(1/\varepsilon^2n))$.
\end{lemma}

\begin{proof}
Any algorithm that solves~$\operatorname{INT}^\varepsilon$ with probability~$>2/3$ can be amplified to fail with probability at most~$\delta$ for any constant~$\delta$ by repetition. We thus assume the algorithm has a small enough failure probability, to be specified later.
Denote by~$\alpha=\alpha(n):=\frac{\log\left({1}/\left({\varepsilon^2 n}\right)\right) - 2}{\log n}$. 
For this choice,~$\varepsilon n < \frac{1}{2}\sqrt{n^{1-\alpha}}$, and thus an algorithm that solves~$\operatorname{INT}^\varepsilon$ can easily be used to solve $\operatorname{INT\text{-}P}^\alpha$.
Note that for~$n^{-3/4}<\varepsilon < 1/(2e^{500}\sqrt{n})$ we have~$\frac{1000}{\ln n} < \alpha < \tfrac12$.
We may ignore the range~$\varepsilon \leq n^{-3/4}$ as for~$\varepsilon = 2n^{-3/4}$ we already we get a lower bound of~$\Omega(n \log n)$ which matches the upper bound even for~$\varepsilon=0$.

By Lemma \ref{lem:design-exists}, there exists a design $\mathcal{X}$ of size $\exp(c\cdot\alpha n \ln n)$. By Lemma \ref{lem:Y-family-describes-X-family}, there exists some set-family $\mathcal{Y}$ of size $k\cdot \alpha n \ln n$ that describes $\mathcal{X}$.
Consider the input distribution in which Alice gets a random set~$X$ from~$\mathcal{X}$ and Bob gets a random set~$Y$ from~$\mathcal{Y}$.
Applying Yao’s minimax principle \cite{yao1977}, it suffices to lower-bound the distributional one-way communication complexity of any deterministic protocol that fails on this distribution with probability less than~$\delta$.

As $|\mathcal X|=\exp(c\alpha n\ln n)$, a uniformly sampled $X\in\mathcal X$ has entropy
\[
  H(X)=\log_2|\mathcal X|
      =\frac{c\alpha n\ln n}{\ln 2},
\]
and for simplicity we modify the constant~$c$ to ``absorb" the factor $\frac{1}{\ln 2}$.
By Lemma \ref{lem:Y-family-describes-X-family}, the sequence $\{\operatorname{INT-P}^\alpha(X, y) : y \in \mathcal{Y}\}$ uniquely determines $X$, therefore
\[
H\left(\{\operatorname{INT-P}^\alpha(X, y)\}_{y \in \mathcal{Y}}\right) \ge H(X)\ge c\cdot \alpha n \ln n.
\]
Let $A(X)$ be the message sent by Alice in a deterministic one-way protocol. Let $\delta$ be small enough that $\delta<1/3$ and $H(\delta) < \frac{c}{2k}$, and let $t = |\mathcal{Y}| = k\alpha n \ln n$. Then,
\begin{align*}
H(\delta) 
&\ge H(\operatorname{INT-P}^\alpha(X,Y) \mid A(X), Y) \quad &\text{by Fano's inequality} \\
&= \frac{1}{t} \sum_{y \in \mathcal{Y}} H(\operatorname{INT-P}^\alpha(X, y) \mid A(X), Y = y) &\\
&= \frac{1}{t} \sum_{y \in \mathcal{Y}} H(\operatorname{INT-P}^\alpha(X, y) \mid A(X)) \quad &\text{since $X \perp Y$} \\
&\ge \frac{1}{t} H\left( \{\operatorname{INT-P}^\alpha(X, y)\}_{y \in \mathcal{Y}} \mid A(X) \right) \quad &\text{by sub-additivity} \\
&= \frac{1}{t} \left( H\left( \{\operatorname{INT-P}^\alpha(X, y)\}_{y \in \mathcal{Y}} \right) - H(A(X)) \right) \quad &\text{by the chain rule} \\
&\ge \frac{1}{t} \left( H\left( \{\operatorname{INT-P}^\alpha(X, y)\}_{y \in \mathcal{Y}} \right) - |A(X)| \right).&
\end{align*}
Rearranging, we have that the message size is at least
\[
|A(X)| \ge H\left( \{\operatorname{INT-P}^\alpha(X, y)\}_{y \in \mathcal{Y}} \right) - t H(\delta).
\]
Since $H\left( \{\operatorname{INT-P}^\alpha(X, y)\}_{y \in \mathcal{Y}} \right) = H(X) \ge c \cdot \alpha n \ln n$ and $t H(\delta) < \frac{c}{2} \alpha n \ln n$, we get,
\[
|A(X)| \ge \frac{c}{2}\cdot \alpha n \ln n = \Omega(\alpha n \ln n)=\Omega(n\log(1/\varepsilon^2n)).
\]
\end{proof}

\subsection{Second Moment Estimation}
\begin{corollary}\label{cor:lower-bound-F2-low-error}
The space complexity of $\left(1\pm \varepsilon\right)$-estimating the second frequency moment of a stream of~$n$ elements from~$U$, for any~$\frac{1}{n^2}\leq \varepsilon \leq \frac{1}{2e^{500}\sqrt{n}}$, is~$$
\Omega\left(n \log \left(\frac{1}{\varepsilon^2 n}\right)\right)
.$$
\end{corollary}
\begin{proof}
Given a streaming algorithm that $(1\pm\varepsilon)$-estimates $F_{2}$ using $s$ bits of memory, we obtain a one-way communication protocol for $\operatorname{INT}^{2\eps}$ that uses $s+O(\log n)$ bits.

Alice feeds her set $X$ (each element once) into the streaming algorithm, obtaining memory image $M$.  She sends $(M,|X|)$ to Bob. Bob resumes the algorithm on his set $Y$, obtains an output $Z$, and returns
\[
m =\frac{Z-|X|-|Y|}{2}.
\]
Let $S$ be the concatenation of $X$ followed by $Y$.  Each element in $X\triangle Y$ has frequency 1 in $S$, each element in $X\cap Y$ frequency 2, so
\[
F_{2}(S)=|X\triangle Y|+4|X\cap Y|
        =|X|+|Y|+2|X\cap Y|.
\]
The above implies that $F_2\le4n$, hence $| Z-F_{2}(S)|\le F_2(S)\eps\le 4 \eps n$, which in turn gives
\[
| m-|X\cap Y||
=\frac12| Z-F_2(S)|
\le2\eps n.
\]
Thus the protocol computes $\operatorname{INT}^{2\eps}$ with the same success probability as the $F_{2}$ estimator (say $2/3$), using $s+O(\log n)$ bits. 
Applying Theorem~\ref{thm:lower-bound-INT-low-error} now yields the claimed
streaming lower bound for $F_{2}$.
\end{proof}

\section{Upper Bounds in the Low Error Regime}
In this section we present matching \emph{upper} bounds to the lower bounds established in the previous section, for both problems of interest: approximate set intersection in the one-way communication model, and second-moment estimation in the streaming model. These results establish the upper bounds required for Theorems~\ref{thm:f2lb} and~\ref{thm:approxset} in the $\varepsilon < 1/\sqrt{n}$ regime.

\subsection{Second Moment Estimation}
\begin{lemma}\label{lem:upper}
    Let~$n$ be large enough,~$1/n^2 \leq \varepsilon \leq 1/\sqrt{n}$, and~$U$ a set of size~$|U|=\text{poly}(n)$. 
    Then, we can~$(1\pm\varepsilon)$-estimate the second frequency moment of a stream of~$n$ elements from~$U$, with probability larger than~$99\%$, using space~$$
    O\left(n \log \left(\frac{1}{\varepsilon^2 n}\right)\right)
    .$$
\end{lemma}

We first observe that any unordered multiset of~$n$ elements from a universe of size~$m$ can be stored using~$\log {n+m-1 \choose n} = O\left(n \log \left(2+\frac{m}{n}\right)\right)$ bits of space.
This immediately implies, for example, the~$O\left(n \log \frac{|U|}{n}\right)$ upper bound when~$|U|\geq2n$.
The algorithm we present to prove Lemma~\ref{lem:upper} is rather natural: we choose~$m\approx \varepsilon^{-2}$ and draw a hash function~$h:U\rightarrow [m]$ that maps our element universe to a smaller one, and then maintain exactly the image of the input stream through~$h$.
Finally, we directly compute the~$F_2$ of the stream.
When~$\varepsilon\approx 1/\sqrt{n}$ and hence~$m\approx n$, and thus, our hash table is expected to contain roughly~$\Theta(n^2/m)\approx\Theta(n)$ hash collisions. The crucial observation is that, as we only care about the number of (actual) collisions in the table, we may subtract from it the expected number of hash collisions and thus vastly reduce the error caused by the hashing. 

\paragraph{Algorithm~$\mathcal{A}$:}
\begin{itemize}
    \item Denote by~$m:=\lceil\frac{201}{\varepsilon^2}\rceil$.
    \item Draw a hash function~$h:U\rightarrow [m]$ from a~$4$-wise independent hash family, and store it in memory.
    \item When receiving the elements of the stream~$S$, maintain in memory the exact multiset~$h(S)$ of images of stream elements through~$h$.
    \item Finally, directly compute~$F:=F_2\left(h\left(S\right)\right)$ the second frequency moment of the maintained multiset, then output~$F':=\frac{m}{m-1}\cdot \left(F-\frac{n^2}{m}\right)$.
\end{itemize}

We note that the space complexity of~$\mathcal{A}$ is as desired, as~$h$ can be stored using~$O(\log n)$ bits of memory and the multiset~$h(S)$ using $O\left(n \log \left(2+\frac{m}{n}\right)\right)$ bits of space.

\begin{claim}
    $\mathbb{E}\left[F'\right]=F_2(S)$.
\end{claim}
\begin{proof}
    Denote by~$x_1,\ldots,x_n$ the elements of the stream~$S$.
    Denote by~$X_{i,j}$ the event that~$h(x_i)=h(x_j)$.
    We have~$F=\sum_{i,j\in [n]} X_{i,j}$.
    We observe that if~$x_i=x_j$ then~$Pr(X_{i,j})=1$, and if~$x_i\neq x_j$ then~$Pr(X_{i,j}=1)=\frac{1}{m}$.
    By linearity of expectation we thus have~$\mathbb{E}\left[F\right]=F_2(S) + \left(n^2-F_2\left(S\right)\right)\cdot \frac{1}{m}$ and in turn~$\mathbb{E}[F']=F_2(S)$.
\end{proof}

\begin{claim}
    $Var(F') \leq \frac{1}{100} \cdot \left(\varepsilon F_2\left(S\right)\right)^2$.
\end{claim}
\begin{proof}
    We begin by bounding~$Var(F)$.
    Denote by~$P:=\{(i,j)\in [n]^2  |  x_i=x_j\}$ the set of~$F_2(S)$ pairs of indices such that the two corresponding stream elements are equal.
    Denote by~${P}^C=[n]^2\setminus P$ all other indices.
    We have
    \begin{align*}
    \mathbb{E}\left[F^2\right] &= 
    \mathbb{E}\left[\left(\sum_{(i,j)\in [n]^2} X_{i,j}\right)^2\right] \\&= 
    \mathbb{E}\left[\left(\sum_{(i,j)\in P} X_{i,j}\right)^2 + 
    2 \left(\sum_{(i,j)\in P} X_{i,j}\right) \left(\sum_{(i,j)\in P^C} X_{i,j}\right) + \left(\sum_{(i,j)\in P^C} X_{i,j}\right)^2
    \right] \\&=
    F_2(S)^2 + 2F_2(S)\cdot \mathbb{E}\left[\sum_{(i,j)\in P^C} X_{i,j}\right] 
    + \mathbb{E}\left[\left(\sum_{(i,j)\in P^C} X_{i,j}\right)^2\right]
    \\&=
    F_2(S)^2 + 2F_2(S)\cdot \left(n^2-F_2(S)\right)\cdot \frac{1}{m} 
    + \mathbb{E}\left[\left(\sum_{(i,j)\in P^C} X_{i,j}\right)^2\right]
    .
    \end{align*}
    Let~$(i,j),(k,\ell)\in P^C$.
    Note that~$x_i\neq x_j$ and~$x_k\neq x_{\ell}$.
    The set~$I:=\{x_i,x_j,x_k,x_\ell \}$ can thus be of size~$2$,~$3$, or~$4$.
    If its size is~$|I|=2$, then~$Pr(X_{i,j}X_{k,\ell})=\frac{1}{m}$. Also, without loss of generality~$x_i=x_k,x_j=x_\ell$ and thus~$(i,k),(j,\ell)\in P$. In particular, there are at most~$2|P|^2$ such pairs.
    If~$|I|\in\{3,4\}$, then~$Pr(X_{i,j}X_{k,\ell})=\frac{1}{m^2}$.
    We therefore have \[
    \mathbb{E}\left[\left(\sum_{(i,j)\in P^C} X_{i,j}\right)^2\right] \leq
    2|P|^2 \cdot \frac{1}{m} + |P^C|^2 \cdot \frac{1}{m^2} =
    2F_2(S)^2\cdot\frac{1}{m} + \left(n^2-F_2\left(S\right)\right)^2 \cdot \frac{1}{m^2}
    .
    \]
    Combining the inequalities above we deduce
    \[
    Var(F) = \mathbb{E}[F^2]-\mathbb{E}[F]^2 \leq
    2F_2(S)^2\cdot\frac{1}{m} \leq
    \frac{2}{201} \cdot \left(\varepsilon F_2\left(S\right)\right)^2
    .
    \]
    We conclude that
    \[
    Var(F')=\left(\frac{m}{m-1}\right)^2 \cdot Var(F) \leq 
    \frac{1}{100} \cdot \left(\varepsilon F_2\left(S\right)\right)^2
    ,
    \]
    for any sufficiently large~$n$ (and in turn~$m$).
\end{proof}

We conclude the proof of Lemma~\ref{lem:upper} by using Chebyshev's inequality.\hfill\qed

\subsection{Approximate Set Intersection}
\begin{corollary}\label{cor:upper-bound-INT-low-error}
For any $\varepsilon<1/\sqrt n$, the randomized one-way communication complexity of $\emph{INT}^\varepsilon$ is $O(n\log(1/\varepsilon^2n))$.
\end{corollary}
\begin{proof}
By the same argument as in Corollary \ref{cor:lower-bound-F2-low-error}, any algorithm that $(1\pm\varepsilon)$-estimates $F_2$ can be converted to a communication algorithm solving $\operatorname{INT}^{2\varepsilon}$ with the same success probability, at an added communication cost of $O(\log n)$, which doesn't change the asymptotic communication bound and implies the desired upper bound.
\end{proof}

\section{Approximate Set Intersection in the High Error Regime}
In this section, we establish a tight $\Theta(1/\varepsilon^2 + \log n)$ bound on the communication complexity of the approximate set intersection problem, for both the one-way and two-way communication models. Combined with the results from the low-error regime, these bounds complete the proof of Theorem~\ref{thm:approxset}.

\subsection{Lower Bound}
\begin{lemma}
    Let $\varepsilon>1/\sqrt{n}$. The randomized communication complexity of $\operatorname{INT}^\varepsilon$ is $\Omega\left(\frac{1}{\varepsilon^2}\right)$.
\end{lemma}
\begin{proof}
    We show a reduction from the Gap Hamming Distance (GHD) communication problem. In GHD, Alice and Bob receive binary strings $x\in\{0,1\}^k$ and $y\in\{0,1\}^k$ respectively, and output (with probability $>2/3$) ``Yes" if $\Delta(x,y) < \frac{1}{2}k-\sqrt{k}$, ``No" if $\Delta(x,y) > \frac{1}{2}k+\sqrt{k}$, and any answer is acceptable if neither condition holds (where $\Delta$ denotes the Hamming distance). Chakrabarti and Regev (\cite{chakrabarti2012GHD}) proved that any randomized protocol solving GHD  requires $\Omega(k)$ bits of communication, even with two-way communication.\\

    Let $k=1/\varepsilon^2$ and let $x,y\in\{0,1\}^k$ be some input to GHD. 
    Assume, without loss of generality, that $k\mid n$ (otherwise pad $n$ up to the next multiple of $k$). We construct two sets $A',B'$ as follows,
    \begin{align*}
    A' &:= \bigcup_{i=0}^{k-1}
            \Bigl\{\bigl(x_i\cdot n + i\cdot\frac{n}{k}+t\bigr)
                   \Bigm| 0\le t<\frac{n}{k}\Bigr\},\\
    B' &:= \bigcup_{i=0}^{k-1}
            \Bigl\{\bigl(y_i\cdot n + i\cdot\frac{n}{k}+t\bigr)
                   \Bigm| 0\le t<\frac{n}{k}\Bigr\}.
    \end{align*}
    Each coordinate $i$ of the GHD strings is “blown up’’ into a block of
    $\frac{n}{k}$ distinct elements.  Hence, $A',B'\subset\left[2n\right],|A'|=|B'|=n$,
    and for every $i$ the two blocks are equal (meaning, they contribute $n/k$ to the intersection size) iff $x_i=y_i$.
    Note that 
    \begin{align*}
    &\Delta(x,y) < \frac{1}{2} k - \sqrt{k} \implies 
    |A' \cap B'| = n - \frac{n}{k} \Delta(x,y) > \frac{1}{2}n + \sqrt{k} \cdot \frac{n}{k} = \frac{1}{2}n + \varepsilon n \quad (\text{``Yes" Instance})\\
    &\Delta(x,y) > \frac{1}{2} k + \sqrt{k} \implies 
    |A' \cap B'| = n - \frac{n}{k} \Delta(x,y) < \frac{1}{2}n - \sqrt{k} \cdot \frac{n}{k} = \frac{1}{2}n - \varepsilon n \quad (\text{``No" Instance}).
    \end{align*}
    
    Let $\Pi(A',B')$ be some communication protocol that that solves $\operatorname{INT}^\varepsilon$.
    The following algorithm will hence solve GHD: Alice calculates $A'$ and Bob calculates $B'$, then they run $\Pi(A',B')$ and answer ``Yes" iff $\Pi(A',B')$ outputs anything larger than $n/2$. As shown above, if the GHD answer is ``Yes", this will indeed be the case, and if the GHD answer is ``No", it will not.
    As the only communication required by this protocol is the one required by $\Pi$, by the above, $\Pi$ requires $\Omega(k)=\Omega(1/\varepsilon^2)$ bits of communication.
\end{proof}

Similarly, the standard reduction from the communication complexity of Equality yields a lower bound of~$\Omega(\log n)$ for private-randomness two-way communication protocols, even when~$\varepsilon$ is constant.

\subsection{Upper Bound}
\begin{lemma}
    Let $\varepsilon>1/\sqrt{n}$. The one-way randomized communication complexity of $\operatorname{INT}^\varepsilon$ is $O\left(\frac{1}{\varepsilon^2}\right)$ in the public-randomness model, and $O\left(\frac{1}{\varepsilon^2} + \log n\right)$ in the private-randomness model.
\end{lemma}

\begin{proof}
We work in the public–randomness model and remark the changes needed for private–randomness at the end.
We present three solutions, each improving on the last.\\

\textbf{Algorithm 1 (One-way, $O((\log n)/\varepsilon^{2})$ bits).}
Define $p=\frac{c}{n\varepsilon^2},c=40$. 
Alice samples $A'\subset A$ such that $\forall a_i\in A:a_i\in A'$ with probability $p$, independently. If $|A'|\ge 10 p|A|$, the algorithm aborts, which by Markov's inequality happens with probability $<0.1$.
Otherwise, Alice sends $(|A|, A')$. Bob outputs $\frac{1}{p} \cdot |A' \cap B|$.\\
\emph{Correctness:}
Denote $C=A \cap B,C'=A' \cap B$, and note that $\forall c\in C: \Pr[c\in C'] = p$. Denote $m=|C|,m'=|C'|$. 
As $\mathbb{E}[m']=mp$, and $\operatorname{Var}[m']\le mp$, by Chebyshev's inequality,
\[
\Pr \left[ \left| \frac{1}{p}m'-m \right|\ge \frac{\varepsilon n}{2}\right] =
\Pr \left[ \left| m'-pm \right|\ge \frac{\varepsilon np}{2}\right] \le 
\frac{\operatorname{Var}[m']}{\varepsilon^2n^2p^2/4} \le 
\frac{4mp}{\varepsilon^2n^2p^2} \le 
\frac{4}{c} \le \frac{1}{10}
\]
By a union bound, the algorithm fails with probability $<0.2$.\\
\emph{Communication cost:} Each element in $A'$ can be written using $O(\log n)$ bits, and $|A|$ is $\log n$ bits long. As $|A'|$ is bounded by $10p|A|\le10c/\varepsilon^2$, the communication cost is $O(\log n / \varepsilon^2)$.

\bigskip
\textbf{Algorithm 2 (Two-way, $O(1/\varepsilon^{2})$ bits).}
Reusing the notations from Algorithm 1, we replace the method for sampling $A'$: Alice and Bob, interpret their shared randomness as a random hash function $h:U\rightarrow [n]$. Using~$h$, they construct the samples,
\[
    A'=\{a\in A:h(a)\le pn\},
    \qquad
    B'=\{b\in B:h(b)\le pn\}.
\]
We note that $C'=A'\cap B=A'\cap B'$ as any element in the intersection is either in both samples or neither. Therefore, as before, $\Pr [c\in C']=p=\Pr [c\in A' \cap B']$.
If $A'$ or $B'$ have size $\ge 10c/\varepsilon^2$, the algorithm aborts, which happens with probability $<0.2$ by the same analysis as above.
After constructing the two sets, Alice and Bob run a two-way communication algorithm for computing $|A' \cap B'|$ exactly using~$O(|A'|+|B'|)$ bits of communication and fails with probability at most~$0.01$, e.g. using the protocol of Brody et al.~\cite{brody2014}. As both sets are of size~$O(1/\varepsilon^2)$, this is also the required communication.
As before, they output $\frac{1}{p}m'=\frac{1}{p}|A'\cap B'|$, which is a correct response with probability $>0.9$. 
Therefore, the algorithm fails with probability $<0.31<1/3$, as claimed.

\bigskip
\textbf{Algorithm 3 (One-way, $O(1/\varepsilon^{2})$ bits).}
We construct $A',B'$ as in Algorithm 2, but replace the two-way intersection sub-routine by the one-way estimator presented in Corollary \ref{cor:lower-bound-F2-low-error}, that computes the same intersection size up to a small additive error:
For $k=\Theta(\frac{1}{\varepsilon^2})$, with $O(k)$ transmitted bits it outputs $z \approx |A'\cap B'|$ up to an additive error $\frac{\varepsilon}{200}k = \Theta(1/\sqrt{k})\cdot k$, and fails with probability $<0.01$.
As Algorithms 1 and 2 gave an error $<\frac{1}{2}\varepsilon n$, the triangle inequality shows we end up with error $<\varepsilon n$ if both Algorithm 2 and the sub-routine succeed. 
The overall failure probability is thus $<0.2+0.1+0.01=0.31<1/3$.

\bigskip
\textbf{Private randomness.}
Newman’s Theorem \cite{newman1991} replaces shared randomness
by private randomness at an additive cost of $O(\log n)$ bits, resulting in
$O(1/\varepsilon^{2}+\log n)$.
\end{proof}

\section{Multipass Histogram Streaming}
In this section we present two-pass streaming algorithm that, with high probability, computes the exact histogram using $O(n\log\log n)$ bits of space. We first present a three–pass algorithm, and then eliminate the third pass while maintaining the same space complexity.
Finally, we consider streaming algorithms with more passes and show that, for exact histogram computation with $r$ passes, $\Theta\!\bigl(n \log^{(\Theta(r))}\! n\bigr)$ bits of memory are both necessary and sufficient.

\subsection{Three-Pass Algorithm}
In this subsection, we present the three-pass algorithm, establishing Theorem~\ref{thm:multipass}. Our algorithm is inspired by the two-way communication protocol of Brody et al.~\cite{brody2014} for computing the exact set-intersection, which repeatedly verifies the equality of large sets and invokes a more expensive local-refinement step only when a set fails verification. This verification of set equality is quite straightforward in the communication setting (each party has full access to its input, so hashing for example is possible), our main contribution is to craft an analogous verification mechanism that operates efficiently in the streaming model.
The algorithm uses the passes, at a high level, as follows.
\begin{enumerate}
    \item Separate the stream elements into buckets by a hash function, and create ``cheap" histograms for each bucket (such that most of them, but not all, are correct).
    \item Verify each of the histograms, to find \emph{all} the ones that are incorrect.
    \item Create full (correct and ``expensive") histograms, in place of those we identified as incorrect.
\end{enumerate}

\begin{lemma}\label{lem:3-pass-algo}
Let $S=\langle x_0,\dots ,x_{T-1}\rangle$ be a stream of $T \le n$ elements from a universe $U$. The memory complexity of calculating the full histogram of $S$ with probability $>1-\frac{3}{n^2}$ using $3$ passes on the stream is $O(n \log\log n)$.
\end{lemma}

\paragraph{Preliminaries:}
We use the following hash functions throughout this section.
\begin{itemize}
    \item $b:U\rightarrow \left[ \frac{n}{\log n} \right]$ is a $6\log n$-wise independent hash function. We refer to all elements with value $j$ in $b$ as the $j$-th \emph{bucket}.
    \item $h:U\rightarrow \mathbb{F}_q$, for some prime $q\in[n^3,2n^3]$, is a  $6\log(n)$-wise independent hash function in which for every element $x\in U$, the value $h(x)$ is uniformly distributed in~$\mathbb{F}_q$.
    \item $\forall j \in \left[ \frac{n}{\log n} \right], g_j:U\rightarrow [2^{10\log\log n}]$ is a pairwise independent hash function. We refer to $g_{b(x_i)}(x_i)$ as the \emph{fingerprint} of $x_i$.
\end{itemize}

\noindent\textbf{The algorithm:}

\textbf{Pass 1: Create a histogram of fingerprints for each bucket.}
For each $j\in \left[ \frac{n}{\log n}\right]$, initialize an empty histogram $H_j$.
The structure of the histograms is a list of pairs, each containing a fingerprint and a counter of appearances of this fingerprint in this bucket. 
The histogram begins empty, and when an element $x_i$ arrives, we check if $g_{b(x_i)}(x_i)$ (the fingerprint of $x_i$) was already encountered in the bucket~$b(x_i)$. If it is new, we add it to the histogram, with counter $1$. If it is not new, we increment the respective counter.
We use the notation $H_j[y]$ for the number of appearances of fingerprint $y$ in the $j$-th bucket (maintained in $H_j$ at the end of pass 1).

\textbf{Pass 2: Verify the different histograms.}
Below, we describe an algorithm \emph{Verify} that verifies assumed histograms.
We use the same hash function $h$ in all calls we make to the \emph{Verify} algorithm.
For each $j\in \left[ \frac{n}{\log n}\right]$, run the \emph{Verify} algorithm on the $j$-th bucket and histogram (we denote $S_j$ for the substream of elements mapped to this bucket).
Let $\mathcal{F} \coloneqq \left\{ j : \emph\{Verify(S_j,H_j)\}=False \right\}$, and refer to the histograms $H_j$ for $j\in \mathcal{F}$ as the failed histograms. If $|\mathcal{F}|\ge\frac{n}{(\log n)^6}$, the algorithm aborts.

\textbf{Pass 3: Correct the failed histograms.}
Delete the failed histograms, and as the stream elements arrive again, create a new histogram $H_j$ for $j\in \mathcal{F}$ with the full names $x_i$ instead of the shorter fingerprints.
Keep the histograms that did not fail unchanged. 

\textbf{Output.}
With high probability, the following holds: For $x_i$ such that $b(x_i)\notin \mathcal{F}$, the number of appearances of $x_i$ in $S$ is $H_{b(x_i)}[g_{b(x_i)}(x_i)]$, and for $x_i$ such that $b(x_i)\in \mathcal{F}$, the number of appearances is $H_{b(x_i)}[x_i]$.

\paragraph{}
We next describe the \emph{Verify} procedure.
\begin{lemma}[One–pass histogram verification]\label{lem:onePassVerify}
Let $S=\langle x_0,\dots ,x_{T-1}\rangle$ be a stream of $T \le n$ elements from the universe $U$, and assume that $S$ contains at most $6\log(n)$ \emph{unique} elements.
Let \emph{fingerprint} be a function that returns a fingerprint for each element in $U$. 
Let $H$ be the histogram of $\langle y_0,\dots ,y_{T-1}\rangle$, where $y_i\coloneqq fingerprint(x_i)$. $H[y_i]$ holds the number of appearances of $y_i$ in $\langle y_0,\dots ,y_{T-1}\rangle$.
Let $h:U\rightarrow \mathbb{F}_q$ be a hash function as described above.
Then, there exists an algorithm \emph{Verify}(S,H) that makes a single pass over the stream $S$, reads $H$ at random access, uses~$O(\log n)$ bits of memory, and satisfies the following:
\begin{itemize}
  \item   
        If all fingerprints in $H$ are distinct
        ($y_i\neq y_j$ for every $x_i\neq x_j$),
        then $\emph{Verify}(S,H)=\text{True}$ with probability $1$.
  \item 
        If some pair $i\neq j$ has the same fingerprint
        $(y_i=y_j$) but different identities
        ($x_i\neq x_j$),
        then $\emph{Verify}(S,H)=\text{False}$
        with probability at least $1-\frac{1}{n^{3}}$ over the choice of $h$.
\end{itemize}
\end{lemma}

\begin{proof}
As $h$ is $6\log n$-wise independent, and $S$ is assumed to have no more than $6\log n$ unique elements, $h$ can be treated as fully uniform and independent on the elements in $S$.
For each index $i$ let $first(x_i)$ be the \emph{first} stream element whose fingerprint equals $fingerprint(x_i)$. 
We define two polynomials in $\mathbb{F}_q[z_1, \dots,z_{|U|}]$:
\[
\text{TrueSum}(z_1, \dots,z_{|U|})=\sum_{i=1}^{|U|} z_{x_i},
\qquad
\text{HistSum}(z_1, \dots,z_{|U|})=\sum_{i=1}^{|U|} z_{first(x_i)}.
\]

If some pair $i<j$ satisfies $x_i\neq x_j$ but $y_i=y_j$, the coefficients of $z_j$ differ in the two polynomials, hence $\text{TrueSum}\not\equiv\text{HistSum}$ (as polynomials); if all fingerprints are unique, the polynomials coincide. The algorithm \emph{Verify} simply evaluates the two polynomials over a random assignment $z_{x_i}\gets h(x_i)$ to their variables, and finally compares the results.

\medskip
\paragraph{Streaming evaluation (One-pass, $O(\log n)$ memory).}
We maintain two accumulator registers, \texttt{accTrue} and \texttt{accHist}. 
\begin{itemize}
    \item To evaluate \text{TrueSum}, for every element $x_i$ we encounter, we add $h(x_i)$ to \texttt{accTrue} ($\text{mod }q$).
    \item To evaluate \text{HistSum}, for every element $x_i$ we encounter, we check if the $fingerprint(x_i)$ was observed before. We can differentiate between ``new" and ``not new" fingerprints by keeping an additional bit next to the fingerprints in the histogram, which doesn't change the asymptotic memory requirements of the histogram. 
    \begin{itemize}
        \item If $fingerprint(x_i)$ is a new fingerprint, we add $\left(h(x_i)\cdot H[fingerprint(x_i)]\right)$ to \texttt{accHist} ($\text{mod }q$).
        \item If $fingerprint(x_i)$ is not a new fingerprint, we ignore it.
    \end{itemize}
\end{itemize}
Note that $\texttt{accTrue}=\text{TrueSum}(h(x_1),\dots,h(x_{|U|}))$ and $\texttt{accHist}=\text{HistSum}(h(x_1),\dots,h(x_{|U|}))$ at the end of the pass.

\medskip
\textbf{Decision.}  Output \emph{True} iff $\texttt{accTrue}=\texttt{accHist}$.

\medskip
\textbf{Error bound.}
If all fingerprints are unique, the test always accepts. Otherwise $\Delta(z)=\text{TrueSum}(z)-\text{HistSum}(z)$ is a
non-zero multi-linear polynomial. By Schwartz–Zippel,
\[
\Pr_{h \in \mathbb{F}_q[z_1,\dots,z_{|U|}]} \bigl[\Delta(h(x_1),\dots,h(x_{|U|}))=0\bigr] \le\frac{\deg\Delta}{q} \le \frac{1}{n^{3}} .
\]
Hence, the probability that the algorithm falsely accepts is bounded by $\frac{1}{n^{3}}$, as claimed.

\medskip
\textbf{Memory cost.} We maintain two registers in $\mathbb{F}_q$, each requires $\log q$ bits to store. Overall, the memory cost is $O(\log q)=O(\log 2n^3)=O(\log n)$.

\end{proof}

The following Lemmas are next used in the correctness proof for the full algorithm.
\begin{lemma} \label{lem:light-primary-bucket}
    Let $X\subseteq U$ such that $|X|\le n$. 
    Let $b:U\rightarrow \left[ \frac{n}{\log n} \right]$ be a hash function as described in the preliminaries. We call $b^{-1}(j)$ the $j$-th bucket.
    With probability at least $1-\frac{1}{n^2}$, every bucket contains at most $6\log n$ distinct elements from $X$.
\end{lemma}

\begin{proof} 
    Fix some bucket $j$. The expected number of distinct elements $x\in X$ in this bucket is at most $\log n$.
    For any $x_i\in X$, let $Z_i$ be an indicator for $b(x_i)=j$. Let $Z\coloneqq \sum_{i=1}^nZ_i$ indicate the number of distinct stream elements in this bucket, and $\mu\coloneqq \mathbb{E}[Z]\le \log n$.
    Schmidt--Siegel’s limited-independence Chernoff bound ~\cite{SchmidtSiegel95} 
    states that the multiplicative Chernoff bound can be applied on binary variables that are $k$-wise independent for $k\ge \lceil \frac{\delta\mu}{1-\mu/n} \rceil$. In our case, as $b$ is $6\log n$-wise independent, the bound is applicable for $\delta=5$ (for large enough $n$), obtaining that the probability that the $j$-th bucket has more than $6\log n$ \emph{distinct} stream elements is bounded by $\exp\left(-\frac{5^2\log n}{7}\right)<\frac{1}{n^3}$.
    A union bound over all buckets implies that the probability that \emph{all} buckets contain at most $6\log n$ elements is at least $1-\frac{n}{\log n} \cdot \frac{1}{n^3} > 1-\frac{1}{n^2}$.
\end{proof}

\begin{lemma} \label{lem:few-fingerprint-collisions}
    Denote $k=\frac{n}{\log n}$, and let $X_1,\dots,X_k\subseteq U$ be sets that are pairwise disjoint, such that $\forall i\in [k],|X_i|\le 6\log n$. 
    Let $g_i:U\rightarrow [2^{10\log\log n}]$ be hash functions as described in the preliminaries. For $x_j\in X_i$ we call $g_i(x_j)$ the \emph{fingerprint} of $x_j$. We say a set $X_i$ contains a fingerprint collision if $\exists x_k\ne x_m \in X_i$ such that $g_i(x_k)=g_i(x_m)$. 
    Then, with probability at least $1-\frac{1}{n^2}$, less than $\frac{n}{(\log n)^6}$ of the sets $X_1,\dots,X_k$ contain a fingerprint collision.
\end{lemma}
\begin{proof}
      For every set $X_i$ let $Z_i$ be an indicator that there is a fingerprint collision in this set, and let $Z\coloneqq \sum_{i=1}^{k} Z_i$.
      As $g_i$ is pairwise independent,
      $
           p:=\Pr[Z_i=1]
             \le\frac{\binom{6\log n}{2}}{(\log n)^{10}}
             \le\frac{18}{(\log n)^{8}}
      $, and by linearity of expectation, $\mu \coloneqq \mathbb{E}[Z]\le\frac{18n}{(\log n)^9}$.
      As the different $g_i$ hash functions are independent of one another, the $Z_i$ are independent random variables, therefore, by a Chernoff bound,
      \[
      \Pr \left[ Z\ge \frac{n}{(\log n)^{6}} \right] \le
      \Pr \left[ Z\ge (1+ (\log n)^2)\mu \right] \le
      \exp\left( -\frac{18n}{(\log n)^3} \right) < \frac{1}{n^2}.
      \]
      Hence, with probability at least $1-\frac1{n^{2}}$, less than $\frac{n}{(\log n)^6}$ sets have a fingerprint collision.
\end{proof}

\paragraph{Correctness of the main algorithm.}

\begin{enumerate}
  \item \textbf{Light primary buckets.}
        By Lemma \ref{lem:light-primary-bucket} applied on $X$ which is the set of unique elements in $S$, the probability that \emph{all} buckets contain at most $6\log n$ \emph{distinct} stream elements is at least $1-\frac{1}{n^2}$.
        
  \item \textbf{Few fingerprint collisions.}
      Conditioned on the event of Item~1, every bucket contains at most $6\log n$ distinct elements.
      By Lemma \ref{lem:few-fingerprint-collisions} applied on $X_1,\dots,X_j$ that are the sets of element in each bucket, with probability at least $1-\frac1{n^{2}}$, less than $\frac{n}{(\log n)^6}$ buckets have a fingerprint collision.

  \item \textbf{Correctness of the histograms.}
        Conditioned on the events of Items 1 and 2, all buckets have at most $6\log n$ elements, and at most $\frac{n}{(\log n)^6}$ buckets contain a fingerprint collision. 
        When a bucket is collision-free, the histogram generated in the first pass is always accurate. Also, any histogram (re)generated in the third pass is also always correct. Therefore, the probability of error is the probability that for some bucket, there exists a fingerprint collision, but \emph{Verify} returned $True$. By Lemma \ref{lem:onePassVerify}, the probability that $\emph{Verify}(S_j,H_j)=True$ for some bucket that has a fingerprint collision is bounded by $2/n^3$. By a union bound, over all buckets with a fingerprint collision, the probability that \emph{Verify} returned this false positive for any such bucket is bounded by $\frac{2}{n^3}\cdot \frac{n}{(\log n)^6} \le \frac{1}{n^2}$.
       
\end{enumerate}
Adding the three failure probabilities gives
\[
  \Pr \bigl[\text{The algorithm outputs \emph{True} but $H$ is wrong}\bigr]
  <\frac{1}{n^{2}}+\frac{1}{n^{2}}+\frac{1}{n^{2}}
  =\frac{3}{n^{2}},
\]
as claimed.

\paragraph{Memory cost.}
\begin{description}
  \item \textbf{Hash functions.}
    \begin{itemize}
      \item $b$ is $6\log n$-wise independent, requiring $O(\log(n)^2)$ bits of memory.
      \item $h$ is $6\log n$-wise independent, requiring $O(\log(n)^2)$ bits of memory.
      \item $g_j$ are $\frac{n}{\log n}$ different pairwise independent functions, each requiring $O(\log n)$ bits of memory, so the overall memory requirement is $O\left( \frac{n\log n}{\log n} \right)=O(n)$.
    \end{itemize}
  Therefore the overall memory cost of the hash functions is $O(n)$.
  
  \item[Counters in histograms.] 
    As the total sum of the counters is $n$, even if we wrote the counters in unary base instead of binary, we would use only $n$ bits. 
  \item[Fingerprints.] For each $j\notin\mathcal{F}$, we save a $10\log\log n$-bit fingerprint, and for each $j\in\mathcal{F}$, we save the full name, in $\log|U|=O(\log n)$ bits of memory.
  \item[Overall cost of fingerprints in histograms.]
     We assume every bucket to have at most, $6\log n$ \emph{distinct} fingerprints. Therefore, the amount of fingerprints for which we save the full $\log n$-bit fingerprint is bounded by $|\mathcal{F}|\cdot 6\log n$. The cost of the rest of the fingerprints is at most, $n \cdot 10\log\log n$. 
     As we assume $|\mathcal{F}|\le \frac{n}{(\log n)^6}$, the overall cost of fingerprints is bounded by $\frac{n}{(\log n)^6} \cdot 6\log n \cdot \log n + n\cdot 10\log\log n=O(n\log\log n)$.
     
  \item[Memory cost of \emph{Verify}.]
     By Lemma \ref{lem:onePassVerify}, the memory cost of \emph{Verify} is $O(\log n)$ per bucket, overall $O(n)$.

\end{description}
Summing all contributions,
\[
   O(n) +O(n) + O(n\log\log n) + O(n)=O(n\log\log n)
\]
bits of total memory, completing the proof.

\subsection{Two–Pass Algorithm}
We now present a two–pass algorithm that eliminates the third pass from Lemma~\ref{lem:3-pass-algo} while preserving the $O(n\log\log n)$ space bound. In the three–pass algorithm, a verification is carried out on the approximate histograms we constructed \emph{for each bucket}: for each bucket, we define two multi-linear polynomials over the variables $\{z_x\}_{x\in U}$, one encoding the true frequencies (\emph{TrueSum}) and the other the frequencies induced by that bucket’s approximate histogram (\emph{HistSum}). Their difference, the bucket's \emph{discrepancy polynomial} $\Delta(z)\coloneqq \text{TrueSum}(z)-\text{HistSum}(z)$, has a variable for each element $x\in U$ with a coefficient equal to the amount by which $x$ is over/under-counted; only elements with fingerprint collisions contribute non-zero coefficients. 
We then verify the correctness of a bucket's histogram by testing whether its discrepancy polynomial is the zero polynomial. 
Our key change is to define a similar single \emph{global} polynomial, rather than one for each bucket, over a much larger field. We then \emph{recover} the coefficients of that global $\Delta$ rather than merely testing if it is the zero polynomial. 
As only a small number of non-zero coefficients appear in~$\Delta$, ideas from sparse recovery allow us to fully reconstruct it with high probability.
We then adjust the affected histogram entries according to the recovered coefficients, yielding the exact histogram without a third pass.

\begin{lemma}\label{lem:2-pass-algo}
Let $S=\langle x_0,\dots ,x_{T-1}\rangle$ be a stream of $T\le n$ elements drawn
from a universe $U$.  
There is a two–pass streaming algorithm that outputs the \emph{exact}
histogram of~$S$ using $O\bigl(n\log\log n\bigr)\text{ bits of memory}$ and succeeds with probability at least $1-\frac{4}{n^{2}}$.
\end{lemma}

\begin{proof}
We reuse the hash functions $b:U\to\bigl[\tfrac{n}{\log n}\bigr]$ and
$\{g_j:U\rightarrow [2^{10\log\log n}]\}_{j\in[n/\log n]}$ from the proof of Lemma~\ref{lem:3-pass-algo}.

\begin{definition}
    Let $\mathcal{D}$ denote the family of multi-linear polynomials $\delta(z_{x_1},\dots,z_{x_{|U|}})$ with at most $6n/(\log n)^5$ non-zero coefficients, where all coefficients are integers in $[-n,n]$. 
\end{definition}

\noindent\textbf{The algorithm.}

\textbf{Pass 1 (fingerprint histograms).}  
Identical to Pass 1 of Lemma~\ref{lem:3-pass-algo}: for every bucket
$j\in[n/\log n]$ we build a list
$H_j=\{(f,c)\}$ of distinct fingerprints
$f=g_j(x)$ together with their counters in the bucket $j$.

\medskip
\textbf{Pass 2 (sparse recovery).}
Choose a prime $q$ with $3^n<q<2\cdot3^n$ and pick a random hash function $h:U\rightarrow\mathbb{F}_q$.

For each index $i$ let $first(x_i)$ be the \emph{first} stream element $x_\ell$ such that $b(x_\ell)=b(x_i)$ and $fingerprint(x_\ell)=fingerprint(x_i)$. 
Define two polynomials in $\mathbb{F}_q\left[z_1, \dots,z_{|U|}\right]$:
\[
\text{TrueSum}\left(z_1, \dots,z_{|U|}\right)=\sum_{i=1}^{|U|} z_{x_i},
\qquad
\text{HistSum}\left(z_1, \dots,z_{|U|}\right)=\sum_{i=1}^{|U|} z_{first(x_i)}.
\]
Let $\Delta\left(z_1, \dots,z_{|U|}\right)\coloneqq\text{TrueSum}-\text{HistSum}$.
Evaluate \text{TrueSum} and \text{HistSum} at the point $(h(x))_{x\in U}$ by maintaining two accumulators, \texttt{accTrue} and \texttt{accHist}, in $\mathbb{F}_q$:
\begin{itemize}
    \item To evaluate \text{TrueSum}, for every element $x_i$ we encounter, we add $h(x_i)$ to \texttt{accTrue} ($\text{mod }q$).
    \item To evaluate \text{HistSum}, for every element $x_i$, check if the $fingerprint(x_i)$ was observed before in the $b(x_i)$-th bucket. We can differentiate between ``new" and ``not new" fingerprints by keeping an additional bit next to the fingerprints in the histogram, which doesn't change the asymptotic memory requirements of the histogram. 
    \begin{itemize}
        \item If $fingerprint(x_i)$ is a new fingerprint in the bucket, we add $\left(h(x_i)\cdot H_{b(x_i)}[fingerprint(x_i)]\right)$ to \texttt{accHist} ($\text{mod }q$).
        \item If $fingerprint(x_i)$ is not a new fingerprint, we ignore it.
    \end{itemize}
\end{itemize}
At the end, set $res\coloneqq \texttt{accTrue}-\texttt{accHist}\in\mathbb F_q$, and note that $res=\Delta\left(h(x_1),\dots,h(x_{|U|})\right)$.

\medskip
\textbf{Output.}
Let $\{\delta_i\}_{i\in [|\mathcal{D}|]}$ be an enumeration of $\mathcal{D}$. For each $i$, check if $\delta_i\bigl(h(x_1),\dots,h(x_{|U|})\bigr)=res$; on the first match set $\delta=\delta_i$ (if no matches are found, abort).
For this $\delta$, and for each $z_{x}$ with a positive coefficient $a_x$, subtract $a_x$ from the histogram entry of $g_{b(x)}(x)$ in bucket $b(x)$; for each $z_{x}$ with a negative coefficient $a_x$, create a new histogram entry (pick an unused fingerprint) with counter $|a_x|$.

The following Lemma will be used in the correctness proof:
\begin{lemma}\label{lem:size-D}
$|\mathcal{D}| \le 2^n$ (for all sufficiently large $n$).
\end{lemma}
\begin{proof}
Let $s\coloneqq 6n/(\log n)^5$. As throughout the paper, we assume $|U|\le n^{c}$ for a constant $c>0$.
Then,
\[
|\mathcal{D}|
\le \sum_{m=0}^{s}\binom{|U|}{m}(2n+1)^m
\le (s+1)\binom{|U|}{s}(2n+1)^s
\le (s+1)\left(\frac{e|U|}{s}\right)^{s} (2n+1)^s.
\]
Using $|U|\le n^c$ and $s=6n/(\log n)^5$,
\[
\log |\mathcal{D}|
\le \log(s{+}1) + s\Bigl(\log(e|U|/s)+\log(2n{+}1)\Bigr) =o(n).
\]
Hence $\log|\mathcal{D}|\le n$ for all sufficiently large $n$, i.e., $|\mathcal{D}|\le 2^n$.
\end{proof}

\noindent\textbf{Correctness.}
By Lemmas~\ref{lem:light-primary-bucket} and~\ref{lem:few-fingerprint-collisions}, with probability $>1-\tfrac{1}{n^2}$, $\Delta$ has at most $s\coloneqq 6n/(\log n)^5$ non-zero coefficients, and all coefficients are integers in $[-n,n]$, hence $\Delta\in\mathcal{D}$. \\
For any $\mathcal{D}\ni\delta\ne\Delta$, by Schwartz-Zippel and the fact that they are both multi-linear,
\[
\Pr_{h \in \mathbb{F}_q[z_1,\dots,z_{|U|}]} 
\left[ \delta(h(x_1),\dots,h(x_{x_|U|}))=\Delta(h(x_1),\dots,h(x_{x_|U|})) \right] \le \frac{1}{q}\le \frac{1}{3^n}.
\]
Therefore, by a union bound over $\mathcal{D}$, using Lemma~\ref{lem:size-D},
\[
\Pr_{h \in \mathbb{F}_q[z_1,\dots,z_{|U|}]} 
\big[\exists\delta\in\mathcal D\setminus\{\Delta\}:\delta(h)=\Delta(h)\big]
\le\frac{|\mathcal D|}{q}\le\left(\tfrac{2}{3}\right)^{n}.
\]
Finally, if $\delta=\Delta$, then $\text{TrueSum}\equiv \Delta+\text{HistSum}$, hence all coefficients match and the correction step yields the exact histogram. Summing the failure probabilities from the two preliminary lemmas ($\le 2/n^2$) and the probability that some $\delta\ne\Delta$ coincides with $\Delta$ at $h$ gives an overall bound of $<3/n^2$.

\medskip
\noindent\textbf{Representing the hash~$h$.}
Representing a fully random $h:U\to\mathbb{F}_q$ naively requires too much memory. Instead, we use Newman’s public–to–private randomness reduction~\cite{newman1991}, observing that it also works for streaming algorithms.
For any problem with at most $N$ possible inputs, there exists a set of $O(\log N)$ values of the shared randomness such that picking uniformly from this set preserves the success probability up to a small additive error, \emph{simultaneously for all inputs}. In particular, it suffices to only maintain in memory an index of one choice within this family, which requires only~$O(\log \log N)$ bits of space.
In our case, $N=|U|^{n}$ bounds the number of possible streams.

\medskip
\noindent\textbf{Memory cost.}
Pass~1 uses $O(n\log\log n)$ bits (as in Lemma~\ref{lem:3-pass-algo}). In Pass~2 we keep two accumulators in $\mathbb F_q$, which costs $\log q=\Theta(n)$ bits, and we can enumerate candidates $\delta\in\mathcal D$ using $O(n)$ bits of memory by reusing this storage for each evaluation at $h$. The switch from public to private randomness contributes only an additional $O\left(\log \log \left(|U|^n\right)\right)=O(\log n)$ bits of memory.
\end{proof}

\subsection{Higher Number of Passes and Further Remarks}

We remark that an algorithm using only $O(n)$ memory cannot compute $F_2$ exactly with probability $>2/3$, let alone the full histogram, with any constant number of passes over the stream. This follows from the $r$-rounds communication lower bound of $\Omega\left(n\log^{(r)} n\right)$ for the set disjointness problem, established by Sağlam and Tardos~\cite{Saglam2013}. By the standard streaming--communication simulation, their result implies a corresponding $\Omega\left(n\log^{(2r-1)} n\right)$ lower bound for $r$-pass streaming algorithms.

On the positive side, the set-intersection communication protocol of Brody et al.~\cite{brody2014}, on which our streaming algorithm is based, extends naturally to $r$ rounds and yields an $O\left(n\log^{\left(\Omega(r)\right)} n\right)$ upper bound in the communication model. In Appendix~\ref{app:rpass} we use the ideas we introduced to translate their protocol into the streaming setting and extended them to obtain an analogous $r$-pass streaming upper bound. Specifically, we show that using~$(2r-1)$-passes, $O\left(n\log^{(r)} n\right)$ bits of memory suffice for exactly computing the histogram.

Finally, while our three-pass algorithm is also efficient in terms of running time, our two-pass algorithm is not. Since our focus is on space complexity, we presented the simplest solution we could find; however, it is likely that efficiency can be achieved using standard polynomial sparse-recovery techniques~\cite{zippel1979probabilistic,ben1988deterministic,kaltofen1988improved}.

\section{Summary and Open Problems}
We provide tight bounds for estimating the second frequency moment $F_2$ of a data stream in the low-error regime $\varepsilon < 1/\sqrt{n}$. Our results fully characterize the one-pass space complexity of $(1 \pm \varepsilon)$-approximating $F_2$ for all values of $\varepsilon$, matching upper and lower bounds up to constant factors. Central to our approach is a precise understanding of the communication complexity of Approximate Set Intersection, which exhibits a sharp phase transition at $\varepsilon = 1/\sqrt{n}$.

We also present a two-pass streaming algorithm that exactly computes a histogram using $O(n \log \log n)$ bits of memory, showing a strict asymptotic separation between one-pass and constant-pass space complexity for $F_2$ estimation when $\varepsilon < 1/\sqrt{n}$. 
This naturally leads to the following question. 
Is there also a separation between one-pass and $O(1)$-pass streaming algorithms for $F_2$ estimation in the regime $\varepsilon > 1/\sqrt{n}$? 
The older $\Omega(1/\varepsilon^2 + \log n)$ space lower bound for~$F_2$ is now known to be sub-optimal for one-pass yet holds also for any constant number of passes. Can either the lower bound or algorithm be improved in these settings?

We also leave as an open problem characterizing the exact trade-off between the number of passes and memory required for exactly computing a stream's histogram.
For example, using two passes we currently have a gap between our~$O(n\log \log n)$ upper bound and the~$\Omega(n \log \log \log n)$ lower bound implied by the communication complexity of Set Intersection.

\subsection*{Acknowledgments}
OZ's research is supported in part by the Israel Science Foundation, Grant No. 1593/24, and by the Blavatnik Family foundation.

\bibliography{main}

\newcommand{\etalchar}[1]{$^{#1}$}
\begin{thebibliography}{BYJKS04}

\bibitem[AGMS99]{alon1999tracking}
Noga Alon, Phillip~B Gibbons, Yossi Matias, and Mario Szegedy.
\newblock Tracking join and self-join sizes in limited storage.
\newblock In {\em Proceedings of the eighteenth ACM SIGMOD-SIGACT-SIGART symposium on Principles of database systems}, pages 10--20, 1999.

\bibitem[AKO11]{andoni2011streaming}
Alexandr Andoni, Robert Krauthgamer, and Krzysztof Onak.
\newblock Streaming algorithms via precision sampling.
\newblock In {\em 2011 IEEE 52nd Annual Symposium on Foundations of Computer Science}, pages 363--372. IEEE, 2011.

\bibitem[AMS96]{alon1996space}
Noga Alon, Yossi Matias, and Mario Szegedy.
\newblock The space complexity of approximating the frequency moments.
\newblock In {\em Proceedings of the twenty-eighth annual ACM symposium on Theory of computing}, pages 20--29, 1996.

\bibitem[And17]{andoni2017high}
Alexandr Andoni.
\newblock High frequency moments via max-stability.
\newblock In {\em 2017 IEEE International Conference on Acoustics, Speech and Signal Processing (ICASSP)}, pages 6364--6368. IEEE, 2017.

\bibitem[BCK{\etalchar{+}}14]{brody2014}
Joshua Brody, Amit Chakrabarti, Ranganath Kondapally, David~P. Woodruff, and Grigory Yaroslavtsev.
\newblock Beyond set disjointness: the communication complexity of finding the intersection.
\newblock In {\em Proceedings of the 2014 ACM Symposium on Principles of Distributed Computing}, PODC '14, page 106–113, New York, NY, USA, 2014. Association for Computing Machinery.

\bibitem[BGKS06]{bhuvanagiri2006simpler}
Lakshminath Bhuvanagiri, Sumit Ganguly, Deepanjan Kesh, and Chandan Saha.
\newblock Simpler algorithm for estimating frequency moments of data streams.
\newblock In {\em Proceedings of the seventeenth annual ACM-SIAM symposium on Discrete algorithm}, pages 708--713, 2006.

\bibitem[BO10]{braverman2010recursive}
Vladimir Braverman and Rafail Ostrovsky.
\newblock Recursive sketching for frequency moments.
\newblock {\em arXiv preprint arXiv:1011.2571}, 2010.

\bibitem[BOT88]{ben1988deterministic}
Michael Ben-Or and Prasoon Tiwari.
\newblock A deterministic algorithm for sparse multivariate polynomial interpolation.
\newblock In {\em Proceedings of the twentieth annual ACM symposium on Theory of computing}, pages 301--309, 1988.

\bibitem[BVWY18]{braverman2018revisiting}
Vladimir Braverman, Emanuele Viola, David~P Woodruff, and Lin~F Yang.
\newblock Revisiting frequency moment estimation in random order streams.
\newblock In {\em 45th International Colloquium on Automata, Languages, and Programming (ICALP 2018)}. Schloss-Dagstuhl-Leibniz Zentrum f{\"u}r Informatik, 2018.

\bibitem[BYJKS04]{bar2004information}
Ziv Bar-Yossef, Thathachar~S Jayram, Ravi Kumar, and D~Sivakumar.
\newblock An information statistics approach to data stream and communication complexity.
\newblock {\em Journal of Computer and System Sciences}, 68(4):702--732, 2004.

\bibitem[BZ25]{braverman2025optimality}
Mark Braverman and Or~Zamir.
\newblock Optimality of frequency moment estimation.
\newblock In {\em Proceedings of the 57th Annual ACM Symposium on Theory of Computing (STOC)}, Prague, Czech Republic, June 2025.

\bibitem[CKS03]{chakrabarti2003near}
Amit Chakrabarti, Subhash Khot, and Xiaodong Sun.
\newblock Near-optimal lower bounds on the multi-party communication complexity of set disjointness.
\newblock In {\em 18th IEEE Annual Conference on Computational Complexity, 2003. Proceedings.}, pages 107--117. IEEE, 2003.

\bibitem[CR12]{chakrabarti2012GHD}
Amit Chakrabarti and Oded Regev.
\newblock An optimal lower bound on the communication complexity of gap-hamming-distance, 2012.

\bibitem[DKS12]{dasgupta2012}
Anirban Dasgupta, Ravi Kumar, and D.~Sivakumar.
\newblock Sparse and lopsided set disjointness via information theory.
\newblock In Anupam Gupta, Klaus Jansen, Jos{\'e} Rolim, and Rocco Servedio, editors, {\em Approximation, Randomization, and Combinatorial Optimization. Algorithms and Techniques}, pages 517--528, Berlin, Heidelberg, 2012. Springer Berlin Heidelberg.

\bibitem[Fel71]{Feller1971}
William Feller.
\newblock {\em An introduction to probability theory and its applications. {V}ol. {II}.}
\newblock Second edition. John Wiley \& Sons Inc., New York, 1971.

\bibitem[Gan11a]{ganguly2011lower}
Sumit Ganguly.
\newblock A lower bound for estimating high moments of a data stream.
\newblock {\em arXiv preprint arXiv:1201.0253}, 2011.

\bibitem[Gan11b]{ganguly2011polynomial}
Sumit Ganguly.
\newblock Polynomial estimators for high frequency moments.
\newblock {\em arXiv preprint arXiv:1104.4552}, 2011.

\bibitem[GGI{\etalchar{+}}02]{gilbert2002fast}
Anna~C Gilbert, Sudipto Guha, Piotr Indyk, Yannis Kotidis, Sivaramakrishnan Muthukrishnan, and Martin~J Strauss.
\newblock Fast, small-space algorithms for approximate histogram maintenance.
\newblock In {\em Proceedings of the thiry-fourth annual ACM symposium on Theory of computing}, pages 389--398, 2002.

\bibitem[Har60]{Harris1960}
T.~E. Harris.
\newblock A lower bound for the critical probability in a certain percolation process.
\newblock {\em Mathematical Proceedings of the Cambridge Philosophical Society}, 56(1):13–20, 1960.

\bibitem[IW05]{indyk2005optimal}
Piotr Indyk and David Woodruff.
\newblock Optimal approximations of the frequency moments of data streams.
\newblock In {\em Proceedings of the thirty-seventh annual ACM symposium on Theory of computing}, pages 202--208, 2005.

\bibitem[JW19]{jayaram2019towards}
Rajesh Jayaram and David~P Woodruff.
\newblock Towards optimal moment estimation in streaming and distributed models.
\newblock In {\em Approximation, Randomization, and Combinatorial Optimization. Algorithms and Techniques (APPROX/RANDOM 2019)}. Schloss-Dagstuhl-Leibniz Zentrum f{\"u}r Informatik, 2019.

\bibitem[KNW10]{kane2010exact}
Daniel~M Kane, Jelani Nelson, and David~P Woodruff.
\newblock On the exact space complexity of sketching and streaming small norms.
\newblock In {\em Proceedings of the twenty-first annual ACM-SIAM symposium on Discrete Algorithms}, pages 1161--1178. SIAM, 2010.

\bibitem[KSZC03]{krishnamurthy2003sketch}
Balachander Krishnamurthy, Subhabrata Sen, Yin Zhang, and Yan Chen.
\newblock Sketch-based change detection: Methods, evaluation, and applications.
\newblock In {\em Proceedings of the 3rd ACM SIGCOMM conference on Internet measurement}, pages 234--247, 2003.

\bibitem[KY88]{kaltofen1988improved}
Erich Kaltofen and Lakshman Yagati.
\newblock Improved sparse multivariate polynomial interpolation algorithms.
\newblock In {\em International Symposium on Symbolic and Algebraic Computation}, pages 467--474. Springer, 1988.

\bibitem[LW13]{li2013tight}
Yi~Li and David~P Woodruff.
\newblock A tight lower bound for high frequency moment estimation with small error.
\newblock In {\em International Workshop on Approximation Algorithms for Combinatorial Optimization}, pages 623--638. Springer, 2013.

\bibitem[Mei18]{Meir2018}
Jeřábek Meir.
\newblock Communication complexity of approximating the size of set intersection, June 2018.
\newblock Theoretical Computer Science, Stack Exchange.

\bibitem[MW10]{monemizadeh20101}
Morteza Monemizadeh and David~P Woodruff.
\newblock 1-pass relative-error lp-sampling with applications.
\newblock In {\em Proceedings of the twenty-first annual ACM-SIAM symposium on Discrete Algorithms}, pages 1143--1160. SIAM, 2010.

\bibitem[New91]{newman1991}
Ilan Newman.
\newblock Private vs. common random bits in communication complexity.
\newblock {\em Information Processing Letters}, 39(2):67--71, 1991.

\bibitem[NY22]{nelson2022optimal}
Jelani Nelson and Huacheng Yu.
\newblock Optimal bounds for approximate counting.
\newblock In {\em Proceedings of the 41st ACM SIGMOD-SIGACT-SIGAI Symposium on Principles of Database Systems}, pages 119--127, 2022.

\bibitem[She10]{Shevtsova2010}
Irina~G. Shevtsova.
\newblock On the absolute constant in the berry--esseen inequality.
\newblock {\em Doklady Mathematics}, 82(3):862--864, 2010.

\bibitem[SSS95]{SchmidtSiegel95}
Jeanette~P. Schmidt, Alan Siegel, and Aravind Srinivasan.
\newblock Chernoff–hoeffding bounds for applications with limited independence.
\newblock {\em SIAM Journal on Discrete Mathematics}, 8(2):223--250, 1995.

\bibitem[ST13]{Saglam2013}
Mert Saglam and Gabor Tardos.
\newblock { On the Communication Complexity of Sparse Set Disjointness and Exists-Equal Problems }.
\newblock In {\em 2013 IEEE 54th Annual Symposium on Foundations of Computer Science (FOCS)}, pages 678--687, Los Alamitos, CA, USA, October 2013. IEEE Computer Society.

\bibitem[Woo04]{woodruff2004optimal}
David~P Woodruff.
\newblock Optimal space lower bounds for all frequency moments.
\newblock In {\em SODA}, volume~4, pages 167--175. Citeseer, 2004.

\bibitem[WZ12]{woodruff2012tight}
David~P Woodruff and Qin Zhang.
\newblock Tight bounds for distributed functional monitoring.
\newblock In {\em Proceedings of the forty-fourth annual ACM symposium on Theory of computing}, pages 941--960, 2012.

\bibitem[WZ21]{woodruff2021separations}
David~P Woodruff and Samson Zhou.
\newblock Separations for estimating large frequency moments on data streams.
\newblock In {\em ICALP}, 2021.

\bibitem[Yao77]{yao1977}
Andrew~C. Yao.
\newblock Probabilistic computations: Toward a unified measure of complexity.
\newblock In {\em Proceedings of the 18th Annual Symposium on Foundations of Computer Science (FOCS)}, pages 222--227. IEEE, 1977.

\bibitem[Zip79]{zippel1979probabilistic}
Richard Zippel.
\newblock Probabilistic algorithms for sparse polynomials.
\newblock In {\em International symposium on symbolic and algebraic manipulation}, pages 216--226. Springer, 1979.

\end{thebibliography}
\bibliographystyle{alpha}

\appendix 

\section{\texorpdfstring{$O(r)$}{O(r)}–Pass Algorithm for Exact Histogram}\label{app:rpass}
For any natural number $r$, we present an algorithm that uses $2r{+}1$ passes on a stream and, with high probability, computes the exact histogram and uses $O(n\log^{(r+1)} (n))$ bits of space. Our algorithm again follows the communication protocol of Brody et al.~\cite{brody2014}, and the modifications we introduced in the paper's body to translate it into a streaming algorithm. 
We create ``cheap" histograms, repeatedly \emph{verify} large aggregates and \emph{rebuild} only where verification fails. The rebuilt histograms use progressively longer fingerprints, so the number of incorrect histograms shrinks quickly with the passes.
We partition the stream elements randomly into buckets, as before, yet now think of them as the leaves of a fixed depth-$r$ tree --- we then repeatedly run verification and correction procedures on subtrees of that tree hierarchy. 
The passes are used as follows.
\begin{enumerate}
  \item \textbf{Initialization (one pass).} Hash the stream into the tree leaves and build a fingerprint histogram for each leaf using short fingerprints. Most leaf histograms are already correct and a small fraction may have collisions.
  \item \textbf{Step $i=0,\dots,r{-}1$ (two passes per step).}
    \begin{itemize}
      \item \emph{Verify.} For every node $v$ at level $i$, we run a certain equality test between (i) the true multiset of items in the subtree of $v$ and (ii) what the current leaf histograms collectively claim. We \emph{flag} subtrees for which the equality test failed or for which the conditions for running it were not satisfied. 
      \item \emph{Rebuild.} For every flagged subtree $v$, we rebuild the histograms of \emph{all} leaves in its subtree, with a longer fingerprint. 
      The histograms in the leaves of non-flagged subtrees are not modified.
    \end{itemize}
\end{enumerate}

Intuitively, each step eliminates nearly all remaining errors. After step $i$ the fraction of incorrect leaf histograms would be polynomial in $1/\log^{(r-i-1)}(n)$, and thus after the final step all leaf histograms are correct with high probability. The total memory cost is bounded w.h.p. by analyzing the fraction of leaf histograms in which we needed to extend the length of the fingerprints to each possible length.

\begin{lemma}
Let $S=\langle x_0,\dots ,x_{T-1}\rangle$ be a stream of $T \le n$ elements from a universe $U$. There is a streaming algorithm that makes $2r{+}1$ passes, outputs the \emph{exact} histogram of $S$ with probability at least $1-\frac{1}{n^4}$, and with high probability, uses $O\bigl(n\log^{(r+1)} n\bigr)$ bits of memory, where $\log^{(k)}$ denotes the $k$-fold iterated logarithm.
\end{lemma}

\begin{proof}
Throughout we write
\[
  \log^{(0)}n := n,\qquad
  \log^{(i+1)}n := \log\bigl(\log^{(i)}n\bigr),\qquad
  L_i := \log^{(i)}n .
\]

\paragraph{Tree structure.}
Let $\mathcal T$ be a rooted tree of depth $r$. For $0\le i\le r$, let $\mathcal{L}_i$ denote the nodes at distance $i$ from the leaves. For $1\le i\le r$, set the degree at level $i$ to $d_i = L_{r-i}/L_{r-i+1}$. Then $|\mathcal{L}_i| = n/L_{r-i}$ and the number of leaves is $|\mathcal{L}_0| = n/L_r$. For a node $v$, let $c(v)$ be its children and $C(v)$ the set of leaves in $v$’s subtree. For $v\in\mathcal{L}_i$, we have $|C(v)| = L_{r-i}/L_r$.

\paragraph{Preliminaries: Hash and field sizes.}
\begin{itemize}
  \item \textbf{Primary separation into tree leaves.} A (random) hash $b:U\to[n/L_r]$ assigns each item to a tree leaf (which we think of as we did of the buckets in the previous algorithms). $b^{-1}(j)$ is the $j$-th leaf.
  \item \textbf{Level–dependent verifier fields.} For each $\ell\in\{0,\dots,r-1\}$ choose a prime $q_\ell\in[L_{r-\ell-1}^{5},2L_{r-\ell-1}^{5}]$ and a (random) hash $h_\ell:U\to\mathbb{F}_{q_\ell}$.
  \item \textbf{Escalating fingerprint hashes.} Randomly draw fingerprint functions $g^{(0)},g^{(1)},\dots,g^{(r)}$ with
  \[
    g^{(0)}:U\to[L_r^{50}],\qquad
    g^{(k)}:U\to[q_{k-1}^{10}]\ \ \text{for all }k=1,\dots,r.
  \]
  For $k\ge 0$ the range size is $\Theta(L_{r-k}^{50})$ and the length of a fingerprint is $\Theta(\log(L_{r-k}^{50}))=\Theta(L_{r-k+1})$ bits.
\end{itemize}

\paragraph{The algorithm.}
\emph{Initial pass (Build leaf histograms).} For every leaf $u$, we build a histogram $H_u=\{(f,c)\}$ for all elements in~$b^{-1}(u)$ using the fingerprint function $f=g^{(0)}(x)$.

For each $i=0,1,\dots,r-1$ we perform two passes: 
\newline
\emph{First pass of step $i$ (Verify subtrees at level $i$).}
For a node $w$, let $M_w=\#\{t:b(x_t)\in C(w)\}$ be the number of stream elements in $C(w)$ with multiplicities (computable, for example, by summing the counters in leaves of $C(w)$).
For every $v\in\mathcal L_i$, independently and simultaneously, compute $\{M_w\}_{w\in c(v)}$ and do as follows. 
\begin{itemize}
\item  If $i=0$ (i.e.,~$v$ is a leaf) and $M_v\ge L_r^5$, mark $v$ as failed and skip the verifier at $v$. We call such a leaf ``heavy".
\item If $i\ge 1$ and some $w\in c(v)$ satisfies $M_w\ge q_{i-1}$, mark $v$ as failed and skip the verifier at $v$. We call such a child ``heavy".
\item Otherwise, run the level-$i$ verifier at $v$ as described below.
\end{itemize}
For each index $j$ let $first(x_j)$ be the \emph{first} stream element $x_\ell$ such that $b(x_\ell)=b(x_j)$ and $fingerprint(x_\ell)=fingerprint(x_j)$. For each $v\in \mathcal{L}_i$ (that were not marked as failed due to heaviness), restrict to items with $b(x_t)\in C(v)$ and define three polynomials in $\mathbb{F}_{q_i}\left[z_1, \dots,z_{|U|}\right]$:
\begin{align*}
\mathrm{TrueSum}_v(z)=&\sum_{t:b(x_t)\in C(v)} z_{x_t},\quad
\mathrm{HistSum}_v(z)=\sum_{t:b(x_t)\in C(v)} z_{first(x_t)},\quad\\
&\Delta_v(z)=\mathrm{TrueSum}_v(z)-\mathrm{HistSum}_v(z)    
\end{align*}

We call $\Delta_v$ the discrepancy polynomial of $v$. Evaluate $\mathrm{TrueSum}_v$ and $\mathrm{HistSum}_v$ at $(h_i(x))_{x\in U}$ using two accumulators $\texttt{accTrue}_v, \texttt{accHist}_v \in\mathbb{F}_{q_i}$:
\begin{itemize}
    \item If $b(x_t)\notin C(v)$, ignore $x_t$.
    \item For $\mathrm{TrueSum}_v$, add $h_i(x_t)$ to $\texttt{accTrue}_v$ (mod $q_i$) for every $x_t$ with $b(x_t)\in C(v)$.
    \item For $\mathrm{HistSum}_v$, when $x_t$ arrives, check whether $fingerprint(x_t)$ was observed before in leaf $b(x_t)$ (track this by a one-bit flag per fingerprint in $H_{b(x_t)}$). If it is new, add $h_i(x_t)\cdot H_{b(x_t)}[fingerprint(x_t)]$ to $\texttt{accHist}_v$ (mod $q_i$); otherwise ignore $x_t$ for $\mathrm{HistSum}_v$.
\end{itemize}
At the end of the pass, $\texttt{accTrue}_v=\mathrm{TrueSum}_v(h(x)_{x\in U})$ and $\texttt{accHist}_v=\mathrm{HistSum}_v(h(x)_{x\in U})$. If $\texttt{accTrue}_v\ne\texttt{accHist}_v$, mark $v$ as failed, and in this case, we say that the verifier rejects $v$. Let $\mathcal F_i\subseteq\mathcal L_i$ be the set of failed nodes (including the ones marked as failed due to heaviness).

\smallskip
\emph{Second pass of step $i$ (Rebuild failed subtrees).}
For each $v\in\mathcal F_i$ and each $u\in C(v)$, rebuild $H_u$ using the longer fingerprint function $g^{(i+1)}$.

\paragraph{Correctness.}
We call a leaf $u\in \mathcal{T}$ correct after step $i$ if there are no fingerprint collisions for the elements in $u$ in the fingerprint version saved after step $i$. We call a node $v\in \mathcal{T}$ correct after step $i$ if all leaves in $C(v)$ are correct after step $i$.
We note that for any $v\in \mathcal{L}_i$, if $v$ is incorrect after step $i-1$, then the discrepancy polynomial of $v$ (at step $i$) is nonzero over the integers, similarly to Lemma~\ref{lem:onePassVerify}.
Consider a node $v\in\mathcal L_{r-1}$ that is incorrect right before the last step, step $r-1$.
We first analyze the probability of the first pass of the last step \emph{not} marking~$v$ as failed:
As the algorithm always marks $v$ as failed if $\exists w\in c(v):M_w\ge q_{r-2}$, we can assume that $\forall w\in c(v):M_w < q_{r-2}$, hence $M_v < \sum_{w\in c(v)}q_{r-2}<q_{r-1}$. Thus, the discrepancy polynomial at level $r-1$ is nonzero over $\mathbb F_{q_{r-1}}$ (it is nonzero as $v$ is incorrect, and all coefficients are in $[-M_v,M_v]\subset (-q_{r-1},q_{r-1})$). By Schwartz–Zippel, the verifier falsely accepts with probability at most $1/q_{r-1}<1/n^5$. 
Second, we analyze the probability that~$v$ is marked failed but the rebuilding of a leaf histogram in its subtree is incorrect.
The algorithm rebuilds the histograms of all leaves $u\in C(v)$ using the fingerprint function $g^{(r)}$ that has a range of size $q_{r-1}^{10}\ge n^{50}$. As $M_u\le n$, the probability that $\exists x\ne y\in S$ such that $b(x)=b(y)=u$ and $g^{(r)}(x)=g^{(r)}(y)$ is at most
\[
\binom{M_u}{2}/q_{r-1}^{10}\le n^2/n^{50} = n^{-48}.
\]
By a union bound over $u\in C(v)$, the probability that $v$ is incorrect after rebuilding is bounded by $1/n^{47}$.
Summing both cases, $v\in \mathcal{L}_{r-1}$ is incorrect after step $r-1$ with probability at most $1/n^5+1/n^{47}<2/n^5$.
By a union bound over all $v\in \mathcal{L}_{r-1}$, the probability that any leaf is incorrect after step $r-1$ is at most $\frac{n}{L_1}\cdot \frac{2}{n^5}< \frac{1}{n^4}$. When this does not happen, all leaf histograms are correct and the algorithm succeeds.

\paragraph{Memory cost.}
We reuse the space between passes. The verification pass at level $i$ maintains two accumulators per $v\in \mathcal{L}_i$, where each accumulator is in $\mathbb{F}_{q_i}$, and hence requires
\[
\Theta \bigl(|\mathcal{L}_i|\cdot \log q_i\bigr) = 
\Theta \Bigl(\frac{n}{L_{r-i}}\cdot \log L_{r-i-1}\Bigr) = 
\Theta \Bigl(\frac{n}{L_{r-i}}\cdot L_{r-i}\Bigr) = 
\Theta(n)
\]
bits.

\emph{Memory cost for histograms.}
Denote by $f(a)$ the frequency of $a$ in the stream, $S^\star=\{x\in U:f(x)>0\}$ the set of unique elements in the stream, and $n^\star=|S^\star|$ their number. For a node $v$ recall that $M_v$ is the number of stream elements mapped to leaves in $C(v)$ with multiplicities, and denote the number of \emph{unique} such stream elements by
\[
D_v=\sum_{a\in S^\star}\mathbf 1\{b(a)\in C(v)\}.
\]

In the first pass, each element $x\in S^\star$ is assigned a fingerprint of length $10L_{r+1}$. Whenever $H_{b(x)}$ is rebuilt at step $i$, its fingerprint is rebuilt with a new fingerprint of length $\log(q_i^{10})=10\log q_i=\Theta(L_{r-i})$. Histogram rebuilding can be triggered by one of the two following events:

\begin{enumerate}
\item \textbf{$v$ has a heavy child, or is a heavy leaf.} For a leaf $u\in \mathcal{T}$, this is the case where $M_u\ge L_r^5$, and for $v\in \mathcal{L}_i$ where $i\ge 1$, the case where some child $w\in c(v)$ is heavy (that is, $M_w\ge q_{i-1}$). 
\item \textbf{Verifier rejection.} If $v\in\mathcal L_i$ is rejected by the level-$i$ verifier (and in particular, $v$ is not a heavy leaf or a parent with a heavy child, as the verifier is not run on these nodes).
\end{enumerate}

We bound the expected total bits used to maintain fingerprints by bounding the size of fingerprints that were constructed due to these two causes separately.

\medskip
\noindent 1. Fix $i\in \{0,\dots,r-1\}$ and $v\in\mathcal{L}_i$, and define
\[
E_v = 
\begin{cases}
    \{M_v\ge L_r^5\}, & i=0,\\
    \{\exists w\in c(v): M_w\ge q_{i-1}\}, & i\ge 1.
\end{cases}
\]
$E_v$ is the event that triggers $v$ to be marked as \emph{failed} without running the verifier.
\begin{itemize}
    \item If $i=0$, by Markov's inequality,
    \[
    \Pr[E_v] = \Pr[M_v\ge L_r^5] \le \frac{\mathbb E[M_v]}{L_r^5} \le \frac{L_{r}}{L_{r}^{5}} = \frac{1}{L_{r}^{4}}.
    \]
    \item If $i\ge 1$, by a union bound over $w\in c(v)$ and Markov's inequality applied to each $M_w$,
    \[
    \Pr[E_v] \le \sum_{w\in c(v)} \frac{\mathbb E[M_w]}{q_{i-1}}
    = \frac{|c(v)|\cdot L_{r-i+1}}{L_{r-i}^{5}}
    = \frac{(L_{r-i}/L_{r-i+1})\cdot L_{r-i+1}}{L_{r-i}^{5}}
    = \frac{1}{L_{r-i}^{4}}.
    \]
\end{itemize}
Therefore, for all $i$, $\Pr(E_v) \le 1/L_{r-i}^{4} $. Moreover $\mathbb E[D_v]\le L_{r-i}$ and $\operatorname{Var}(D_v)\le \mathbb E[D_v]$ (as $D_v$ is a sum of independent Bernoulli variables), hence for large $n$, $\mathbb E[D_v^2] \le \mathbb E[D_v]^2+\operatorname{Var}(D_v) \le  2L_{r-i}^2$. By Cauchy–Schwarz,
\[
\mathbb E\big[D_v\,\mathbf 1\{E_v\}\big]
\le \big(\mathbb E[D_v^2]\big)^{1/2}\,\Pr(E_v)^{1/2}
\le \sqrt{2}\,L_{r-i}\cdot \frac{1}{L_{r-i}^{2}}
= O\!\left(\frac{1}{L_{r-i}}\right).
\]
There are $|\mathcal L_i|=n/L_{r-i}$ nodes at level $i$, so the expected number of element fingerprints rebuilt due to $E_v$ at level $i$ is
\[
\sum_{v\in\mathcal L_i} \mathbb E\big[D_v\,\mathbf 1\{E_v\}\big]
= O\!\left(\frac{n}{L_{r-i}^{2}}\right).
\]
Each such element receives a fingerprint of length $\Theta(L_{r-i})$ bits, so the contribution of this case at level $i$ is
\[
O\!\left(\frac{n}{L_{r-i}^{2}}\right)\cdot \Theta(L_{r-i}) = O\!\left(\frac{n}{L_{r-i}}\right).
\]
Summing over $i=0,\dots,r-1$ we get
\[
\sum_{i=0}^{r-1} O\!\left(\frac{n}{L_{r-i}}\right)
= O\!\left(\frac{n}{L_r}\right)\!\left(1+\sum_{j=1}^{r-1}\frac{L_r}{L_{r-j}}\right)
= O\!\left(\frac{n}{L_r}\right).
\]
Thus the expected memory used by fingerprints created due to this case is $O(n/L_r)=o(n)$.

\medskip
\noindent 2. Let $x\in S^\star$ be a stream element, we analyze the expected total size of fingerprints that are constructed for it throughout the algorithm due to verification rejections.
Throughout the proof, we condition on a fixed assignment to leaves $b$, and all probabilities are taken over the hash functions $h,\{g^{(i)}\}_{i\in \{0,\dots,r\}}$.
Fix a level $i$, and let $u_x:=b(x), \;v:=p_i(u_x)$.

\emph{For $i=0$:}
Here $v=p_0(u_x)=u_x$ is a leaf. Conditioned on $\neg E_0(v)$, meaning, $M_v<L_r^{5}$ (and in particular $D_v\le M_v<L_r^{5}$), the fingerprint of $x$ is rebuilt at step $0$ if the verifier rejects $v$, which only happens when $H_v$ is incorrect before step $0$. Hence, the probability of this is bounded by the probability that the elements in $v$ have a fingerprint collision under $g^{(0)}$, which is bounded by
\[
\binom{D_v}{2}/L_r^{50} \le 
\binom{L_r^5}{2}/L_r^{50} = 
O(L_r^{-40}).
\]
Therefore, the expected new fingerprint length for $x$ at step $0$ is $\Theta(L_{r})\cdot O\!\left(L_r^{-40}\right) = O\!\left(L_r^{-39}\right)$.

\smallskip
\emph{For $i\ge 1$:}
Conditioned on $\neg E_v$, the fingerprint of $x$ is rebuilt at step $i$ when $v$ is rejected by the verifier at step $i$, and in particular, at least one of its children $w \in c(v)$ is incorrect after step $i-1$. 
We first bound the probability of an incorrect~$w$ being falsely accepted by the step~$(i-1)$ verifier; unless that happens, it is marked and is thus rebuilt at step~$(i-1)$ --- we then bound the probability of that rebuilding leading to an incorrect histogram. 
As the algorithm always marks $w$ as failed if $E_w$ happens, we can assume that $M_w<q_{i-1}$. Indeed, if $w\in \mathcal{L}_0$ this is the definition of $E_w$, and if $w\in \mathcal{L}_i$ for $i\ge 1$, all its children $z\in c(w)$ have $M_z < q_{i-2}$, thus $M_w< \sum_{z\in c(w)}q_{i-2}<q_{i-1}$. Therefore, if $w$ is incorrect before step $i-1$, the discrepancy polynomial of $w$ (at step $(i-1)$) is nonzero modulo $q_{i-1}$ (it is nonzero as $w$ is incorrect, and all coefficients are in $[-M_w,M_w]$). Therefore, by Schwartz–Zippel, the verifier falsely accepts with probability at most $1/q_{i-1}\le 1/L_{r-i}^{5}. $ 
If~$w$ is indeed marked, then every $u\in C(w)$ is rebuilt at step $i-1$ with fingerprints drawn from a range of size $q_{i-1}^{10}$. For each $u\in C(w)$ we have $D_u\le M_u\le M_w<q_{i-1}$, so $u$ is incorrect after this step with probability at most
\[
\frac{\binom{D_u}{2}}{q_{i-1}^{10}}
\le \frac{q_{i-1}^{2}}{q_{i-1}^{10}}
= q_{i-1}^{-8}
\le \frac{1}{L_{r-i}^{8}}.
\]
Thus
\[
\Pr[w\text{ is incorrect after step }(i-1)]
\le \frac{1}{q_{i-1}} + |C(w)|\cdot \frac{1}{L_{r-i}^{8}}
= \frac{1}{L_{r-i}^{5}} + \frac{L_{r-i+1}}{L_r\,L_{r-i}^{8}}.
\]
A union bound over the $|c(v)|=L_{r-i}/L_{r-i+1}$ children of~$v$ yields that the verifier rejects $v$ with probability at most
\[
\frac{L_{r-i}}{L_{r-i+1}}\left(\frac{1}{L_{r-i}^{5}} + \frac{L_{r-i+1}}{L_r\,L_{r-i}^{8}}\right)
= \frac{1}{L_{r-i+1}\,L_{r-i}^{4}} + \frac{1}{L_r\,L_{r-i}^{7}}
\]
(conditioned on $\neg E_v$).
After a rejection at level $i$, the length of the fingerprint of $x$ is increased to $\Theta(L_{r-i})$ bits. Therefore, the expected cost of rebuilding the fingerprint of $x$ at step $i$ is at most
\[
\Theta(L_{r-i})\left(\frac{1}{L_{r-i+1}\,L_{r-i}^{4}} + \frac{1}{L_r\,L_{r-i}^{7}}\right)
= O\!\left(\frac{1}{L_r\,L_{r-i}^{6}}\right)\qquad(i\ge1).
\]
Summing over all levels $i=0,\dots,r-1$ we get
\[
O\!\left(L_r^{-39}\right) \;+\; \sum_{i=1}^{r-1} O\!\left(\frac{1}{L_rL_i^{6}}\right)
= o(1).
\]
Hence the total expected memory cost from verifier rejections is $o(n)$.

\medskip
As both cases above are expected to cost $o(n)$ bits of memory, the probability that they cost more than $cnL_{r+1}$ for some constant $c$ is bounded by 
\[
\frac{o(n)}{c\,nL_{r+1}}
= o\!\left(\frac{1}{L_{r+1}}\right)
\]
by Markov's inequality. As the cost of the initial fingerprints is $O(nL_{r+1})$, and the cost counters kept in the histograms is $O(n)$ (deterministically for both), with probability $>1-1/L_{r+1}$, the total  cost of the histograms is $O(nL_{r+1})$.

\medskip
Finally, we address the memory cost of the hash functions. We assumed the hash functions $b,\{h_i\}_{i\in[r]}$ to be fully independent. Representing them naively would require too much memory. As in the end of Lemma~\ref{lem:2-pass-algo}, we apply a Newman-style reduction~\cite{newman1991}, using only $O(\log n)$ additional bits overall.

\end{proof}

\end{document}